\documentclass[a4paper,fleqn,usenatbib,useAMS]{mnras}

\usepackage[T1]{fontenc}
\usepackage{ae,aecompl}

\usepackage{graphicx}	
\usepackage{amsmath}	
\usepackage{amssymb}	
\usepackage{threeparttable}
\usepackage{tabularx}
\usepackage{booktabs}
\usepackage{xfrac}

\title[Radiative Transfer of HCN]
  {Radiative Transfer of HCN: Interpreting observations of hyperfine anomalies}

\author[Mullins et al.]
{A.M.~Mullins$^1$
\thanks{a.mullins4@nuigalway.ie (A.M.~Mullins)},
R.M.~Loughnane$^{1, 2}$\thanks{r.loughnane@crya.unam.mx (R. M. Loughnane)}, M.P.~Redman$^1$, B.~Wiles$^1$, 
\newauthor N.~Guegan$^1$, J.~Barrett$^1$\,\& E.R.~Keto$^{3,4}$ \\
$^1$Centre for Astronomy, School of Physics, National University of Ireland, Galway.\\
$^2$Insitiuto de Radiostronom\'ia y Astrof\'isica, UNAM, Apartado Postal 72-3 (Xangari), 58089 Morelia, Mexico.\\
$^3$Harvard-Smithsonian Center for Astrophysics, 160 Garden Street, Cambridge, MA 02420, USA.\\
$^4$Max-Planck Institut f{\"u}r Astronomie, K{\"o}nigstuhl 17, D-69117, Heidelberg, Germany.\\
}

\date{Accepted 2016 April 8. Received 2016 April 7; in original form 2014 December 23}

\pubyear{2016}

\begin{document}
\label{firstpage}
\pagerange{\pageref{firstpage}--\pageref{lastpage}}
\maketitle

\begin{abstract}
Molecules with hyperfine splitting of their rotational line spectra are useful probes of optical depth, via the relative line strengths of their hyperfine components.The hyperfine splitting is particularly advantageous in interpreting the physical conditions of the emitting gas because with a second rotational transition, both gas density and temperature can be derived. For HCN however, the relative strengths of the hyperfine lines are {\it anomalous}. They appear in ratios which can vary significantly from source to source, and are inconsistent with local thermodynamic equilibrium. This is the HCN hyperfine anomaly, and it prevents the use of simple LTE models of HCN emission to derive reliable optical depths. In this paper we demonstrate how to model HCN hyperfine line emission, and derive accurate line ratios, spectral line shapes and optical depths. We show that by carrying out radiative transfer calculations over each hyperfine level individually, as opposed to summing them over each rotational level, the anomalous hyperfine emission emerges naturally. To do this requires not only accurate radiative rates between hyperfine states, but also accurate collisional rates. We investigate the effects of different sets of hyperfine collisional rates, derived via the Òproportional methodÓ and through direct recoupling calculations. Through an extensive parameter sweep over typical low mass star forming conditions, we show the HCN line ratios to be highly variable to optical depth. We also reproduce an observed effect whereby the red-blue asymmetry of the hyperfine lines (an infall signature) switches sense within a single rotational transition. 

\end{abstract}



\section{Introduction}

HCN appears an attractive choice of tracer species for molecular gas for a number of reasons: its chemistry has been well studied over a wide range of conditions ({\it e.g.} \citet{Graedel1982,Forets1990,Hebrard2012}), it has a high critical density, and it is relatively abundant. It is routinely used in studies of low mass star forming cores \citep{Sohn07,Daniel2013a}, disks \citep{vanderPlas2014} and high mass star formation \citep{Rolffs2011,Jin2015}. HCN has also been used to observe comets \citep{Hogerheijde2009,Friedel2005,hirota99}, evolved stellar atmospheres \citep{Schilke2003,Smith2014}, active galaxies \citep{Aalto2012,Salas2014} and high redshift molecular clouds \citep{Gao2004}.

\begin{figure*}
	\includegraphics[width=0.4\textwidth]{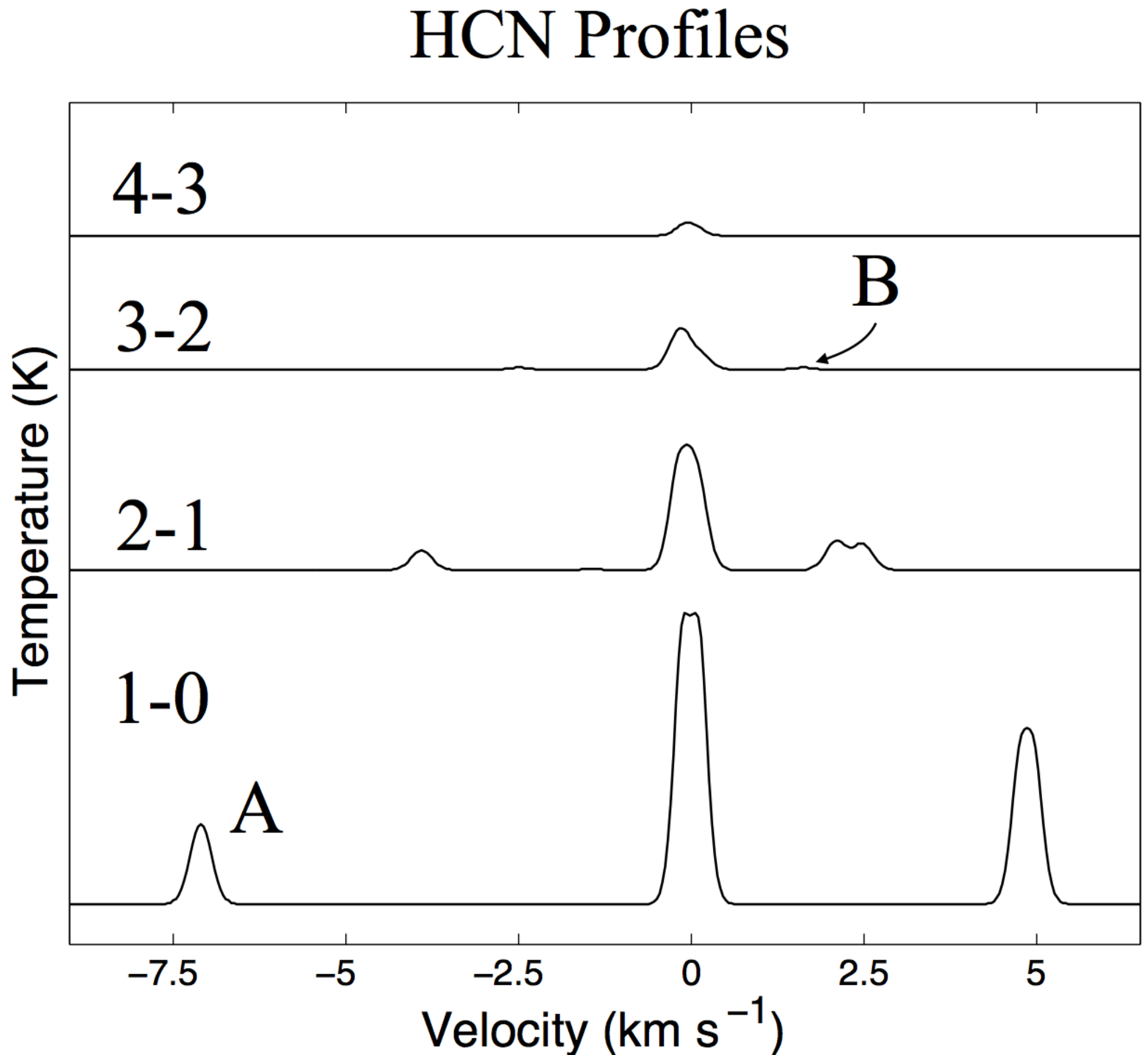}
\quad
\includegraphics[width=0.4\textwidth]{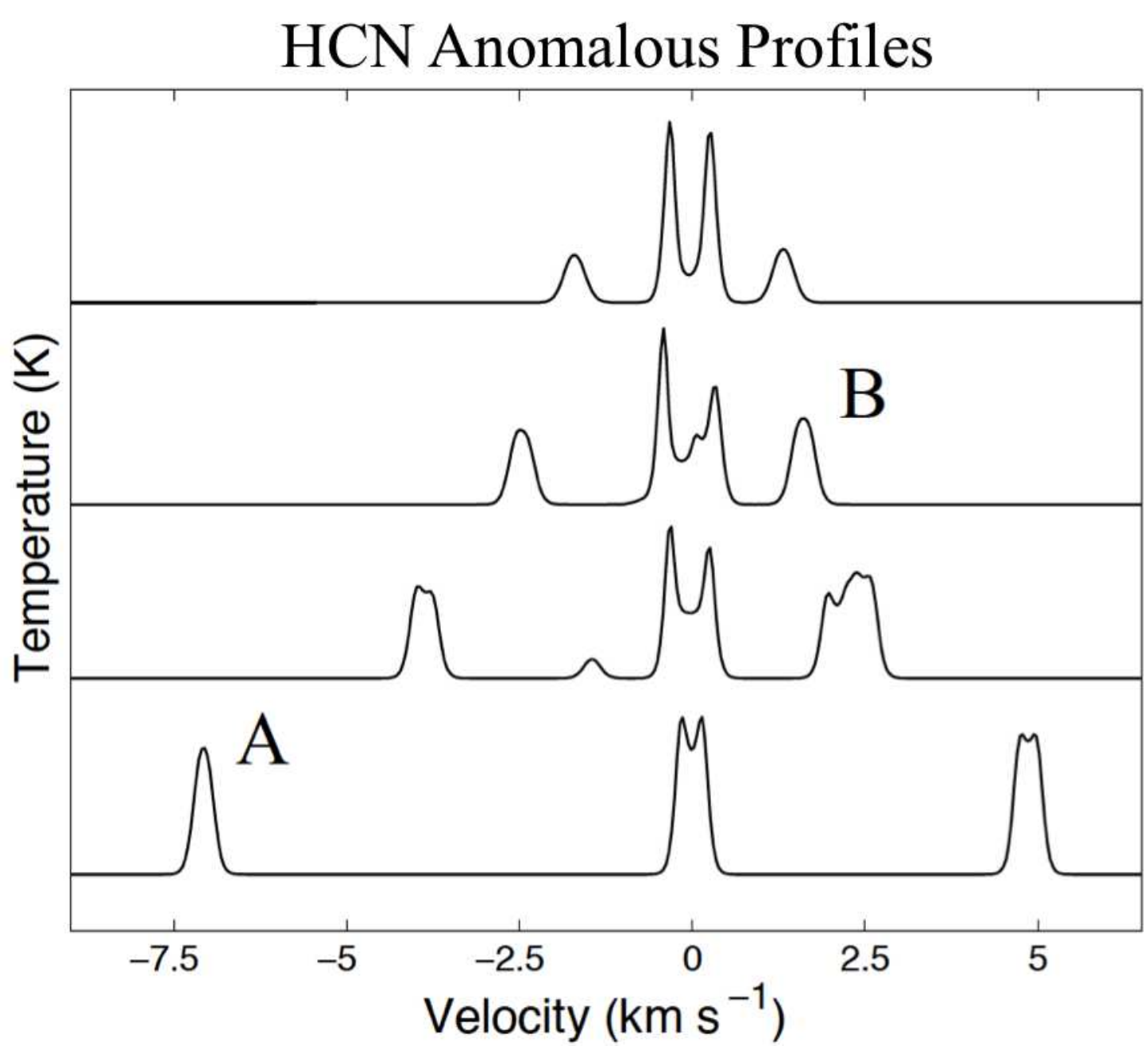}\\~\\

\includegraphics[width=0.42\textwidth]{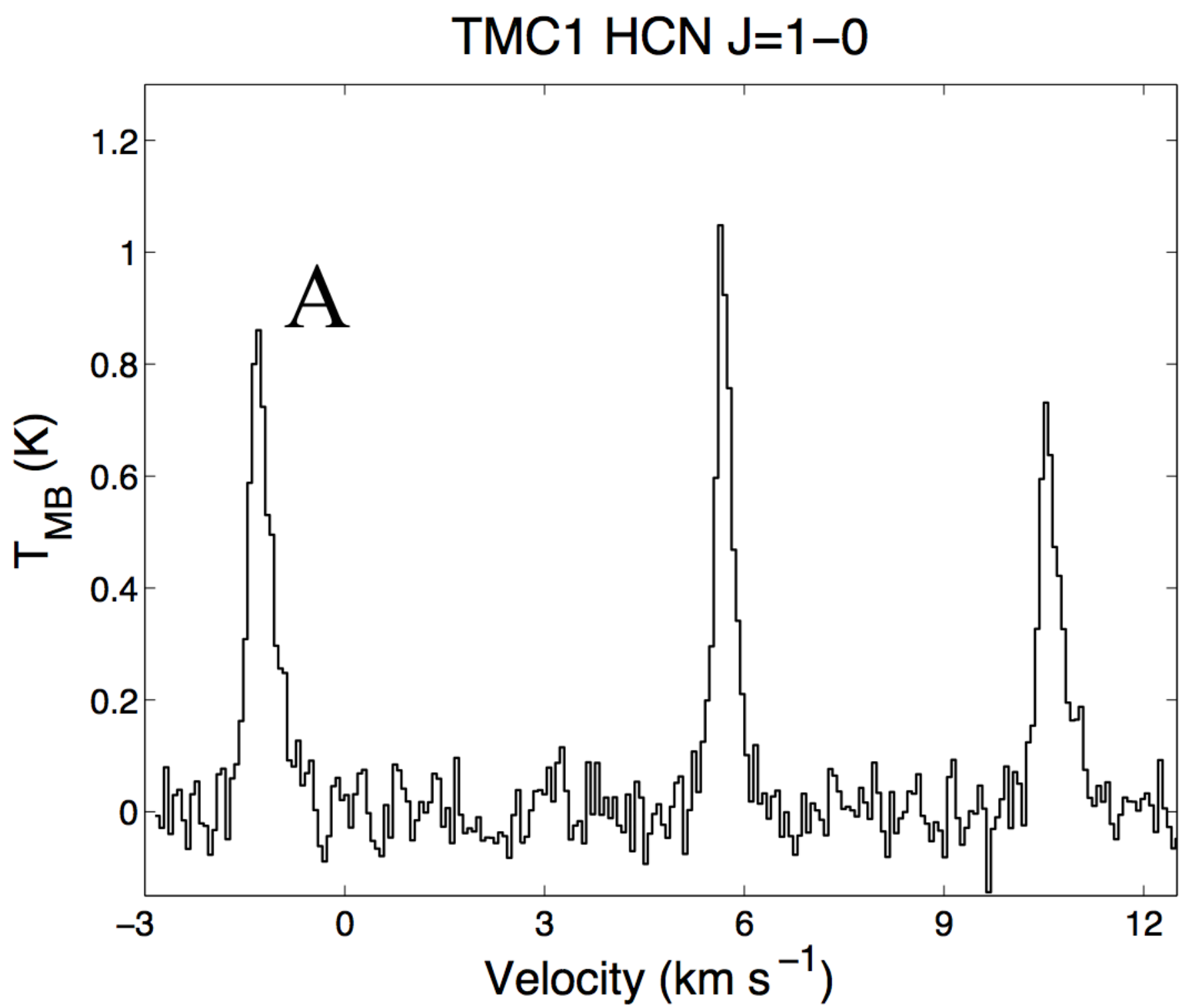}\,
\includegraphics[width=0.41\textwidth]{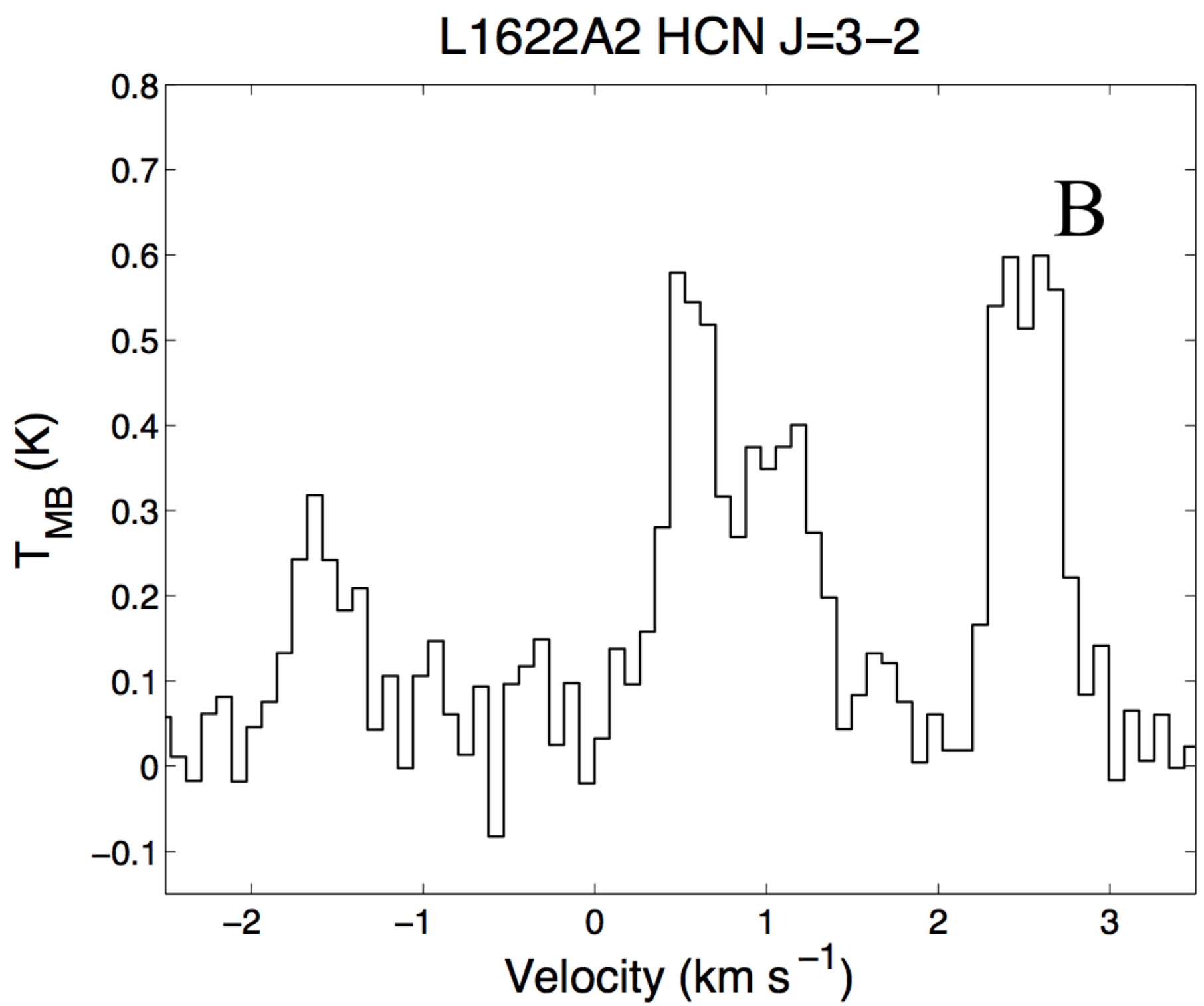}
   \caption{\emph{(top left)} Radiative transfer model of the lowest four rotational lines of HCN. (from top to bottom) J=4$\to$3, J=3$\to$2, J=2$\to$1, and J=1$\to$0, showing the expected LTE line strength ratios between the hyperfine components. \emph{(top right)} The same lines, but from an optically thick, non-LTE model in which the hyperfine line strengths are anomalous. The labels A and B denote the J$_{\rm F}$=$1_{\rm0}\rightarrow0_{\rm1}$, and J$_{\rm F}$=$3_{\rm2}\rightarrow2_{\rm2}$ transitions, which are susceptible to anomalous hyperfine line strengths. \emph{(bottom left)} Observation of TMC-1, with the J$_{\rm F}$=$1_{\rm0}\rightarrow0_{\rm1}$ transition anomalously boosted. \emph{(bottom right)} JCMT observation of L1622A2 with the anomalously boosted transition J$_{\rm F}$=$3_{\rm2}\rightarrow2_{\rm2}$ marked B.}
   	\label{HCN}
   \end{figure*}

The end nitrogen atom of HCN is responsible for its distinct hyperfine structure. The large quadrupole moment leads to widely spaced hyperfine lines, especially for the lowest rotational transitions \citep{Walms82, Sohn07, Loughnane2012a}. Observations of the HCN hyperfine lines have shown that for a given rotational transition, particularly J=1$\rightarrow$0, they are commonly found in ratios that cannot be reproduced by a single excitation temperature model, for any optical depth. In other words, they are prone to anomalous line strengths in their hyperfine spectra \citep{Guill81,Loughnane2012a} (see Figure~\ref{HCN}). 

The degree of anomaly in the J=1-0 transition of HCN can be characterised by the relative strengths of the individual hyperfine lines using the ratios R$_{02}$ and R$_{12}$ \citep{Cern84,Loughnane2012a} where, 
\begin{equation}
\rm{R}_{02} = \frac{T_{(F = 0\to1)}}{T_{(F = 2\to1)}} ;\quad \rm{R}_{12}=\frac{T_{(F = 1\to1)}}{T_{(F = 2\to1)}}
\label{r02}
\end{equation}
The variation in R$_{02}$ (marked \emph{A} in Figure~\ref{HCN}) is particularly exaggerated.  The F=0$\rightarrow$1 transition can sometimes exceed the other lines in strength, which could never occur in a single excitation temperature model, where the values of R$_{02}$ and R$_{12}$ are 0.2 and 0.6 respectively. HCN hyperfine anomalies are widespread among star-forming cores (Pirogov 1999; Sohn et al. 2007; Loughnane et al. 2012), but are particularly evident in low-mass clouds where the ratios in Eq.~\ref{r02} can exceed 1.0 (e.g. Oph D, L1696B: $R_{\rm02}>1.0$, L694-2: $R_{\rm02}>1.0$, $R_{\rm12}>1.0$).   
Many theories have been proposed to explain the anomalies, ranging from photon trapping caused by hyperfine components with differing optical depth \citep{Kwan75}, to scattering of radiation from the core in a moderately dense envelope \citep{Cern84}, to small scale clumpiness in low mass cores with thermal local line widths \citep{Pirog99a}. An intriguing feature of the anomalous intensity problem is that C$^{17}$O, which also exhibits a hyperfine pattern, is not subject to anomalies and agrees with the LTE ratio irrespective of the brightness temperature or optical depth.  \citet{Redman2002}  treat the hyperfine structure of C$^{17}$O as a modification of the rotational lineshape function $\rm{\phi(\nu)}$, and perform the radiative transfer at the J-level only. This approximation will not work for HCN however, as an analysis of the selection rules for electric dipole transitions ($\Delta$J=$\pm$1, $\Delta$F= 0,$\pm$1) between J and F levels up to J = 7 in HCN reveals that there is only a single pathway ($\Delta$J=-1, $\Delta$F= +1) leading to the J$_{\rm F}$ = $1_{\rm 0}\rightarrow0_{\rm1}$ transition out of 36 total pathways to the J=1 level (six transitions between each pair of rotational levels). This leaves the transition very susceptible to line overlap effects in transitions between higher J-levels, as any photons gained or lost are significant compared to the total photon flux into this transition. These overlap effects prevent a single lineshape function being used at the J-level, and force us to consider radiative transfer of each individual hyperfine F-level line. These results are mirrored by those of \citet{Keto10} where the approximation of statistical equilibrium among the hyperfine levels of each rotational level was also found unsuitable to model the hyperfine spectra of N$_2$H$^+$. 

Further investigation shows that the hyperfine lines are found to vary in width within a single rotational line pattern, and for lines exhibiting dynamical effects such as asymmetric profiles (due to doppler broadening in moderate opacity gas), the sense of the asymmetry can switch from red to blue in different hyperfine lines of a single rotational transition \citep{Sohn07, Loughnane2012a}. These behaviours are all impossible to reproduce when HCN is analysed in terms of its rotational structure only, and suggests the necessity of modelling the radiative transfer of each hyperfine line individually.

Here we describe an F-level radiative transfer calculation that is then used to generate J-level emission line spectra for comparison with observed HCN line data. While the line frequencies are measured accurately, and radiative transition rates are well known analytically, determining collisional rate coefficients is a challenging numerical problem. There are different collisional rates available in the literature, involving different approximations. We carried out our radiative transfer calculations using three different sets of rate coefficients consisting either of a set of J-level coefficients (\citet{GT74} as updated for the LAMBDA database \citep{Schoier2005}, \citet{Vera14}), or explicit calculations of the F-level coefficients \citep{BA12}. \citet{Vera14} suggests that the approximate nature of the potential energy surface (PES) for the HCN-H$_2$ system used in \citet{BA12} does not fully account for the effects of the orientation of the H$_2$ molecules on the collisional rates. Thus the collisional rates derived using this approximate PES, including the individual hyperfines or F-level rates, may not be as accurate as rates derived from more accurate calculations of the PES, even if the F-level rates are derived with further approximate methods rather than calculated explicitly. For example, we can scale the J-level rate coefficients derived in \citet{Vera14} into F-level rates by assuming that the J-level rates are scaled into the same proportions as the F-level rates calculated explicitly by \citet{BA12}. For comparison, we also scale the J-level rate coefficients of \citet{GT74} according to the `proportional method', as initially suggested by \citet{Guill81}, and further demonstrated in modelling non-LTE hyperfine line emission for N$_2$H$^+$ in \citet{Keto10}. We find however, the results of our parameter sweep (\S 3.3) are largely unaffected by the choice of rates, in agreement with the conclusions of \citet{Keto10}.

This paper is organised as follows: \S2 describes the treatment of the molecular physics of HCN where we present the formalism employed to account for the hyperfine state-to-state transitions for HCN in our models. In particular, the choice of collisional rate coefficients is discussed in detail. \S3 presents our modeling of the anomalies, with a radiative transfer code (\textsc{mollie}). The model reproduces a spectrum of the core TMC-1, one of the first sources in which the HCN anomalies were observed \citep{Walms82}. An analysis of red-blue asymmetry switching in double-peaked line profiles from star forming cores, is presented and shown also to be reproducible. Finally, a comprehensive parameter sweep investigating the conditions triggering the HCN anomalies is carried out. \S4 then outlines our conclusions and some advice for observers on the use of HCN as a dynamical tracer in star forming regions as well as in general astrophysical conditions.

\section{Molecular Physics of HCN}

In order to describe the physical properties of the molecular cloud gas, we must fully take into account the motion of the gas at the molecular level. We must also consider the relative populations for those levels that are being populated or depopulated as a consequence of the gas dynamics. There are three requirements for a complete treatment:

\begin{enumerate}
\item The frequencies of each individual hyperfine transition.
\item The radiative excitation/de-excitation coefficients, for both spontaneous and stimulated emission/absorption, namely the Einstein A and B coefficients.
\item The collisional rates with both \emph{ortho}-H${\rm_2}$, \emph{para}-H${\rm_2}$, He and electrons. 
\end{enumerate}

\subsection{Line Frequencies and Strengths}
We use the most precise set of frequencies for the rotational and hyperfine levels of HCN to date, based on the calculation of spectroscopic constants by \citet{Ahrens02}. In that work, they improved upon the values determined from molecular beam maser measurements by \citet{MolHCN}, the previously accepted standard for HCN. The results of the hyperfine frequency calculations up to J=5-4 are reproduced in Table~\ref{tab:molspec}.

\subsection{Einstein A and B coefficients}
\begin{table}
\begin{center}
\caption{Spectroscopic values for hyperfine components of the lowest 5 downward J-level transitions, listed in ascending frequency. Here J is the upper and J$^{\prime}$ the lower level. Calculated by the authors based on data from \citet{Ahrens02}.}
\begin{threeparttable}
\begin{tabular}{@{}cccrc@{}}
\hline \hline
&& Frequency & $\rm A_{JF\rightarrow J^{\prime}F^{\prime}}$ &\\[-1ex]
\raisebox{1.5ex}{J F} & \raisebox{1.5ex}{$\rm J^{\prime} F^{\prime}$} & (GHz) & (1$\rm0^{-5}$s$^{-1}$) & \raisebox{1.5ex}{S$_{JF\rightarrow J^{\prime}F^{\prime}}$}\\ [-0.5ex] 
\midrule
1 1 & 0 1 & 88.630413 & 2.405060 & 0.3333\\
1 2 & 0 1 & 88.631846 & 2.405177 & 0.5555\\
1 0 & 0 1 & 88.633935 & 2.405347 & 0.1111\\
 & & & & \\
2 2 & 1 2 & 177.259676 & 5.772518 & 0.0833\\
2 1 & 1 0 & 177.259921 & 12.826738 & 0.1111\\
2 2 & 1 1 & 177.261109 & 17.317900 & 0.2500\\
2 3 & 1 2 & 177.261220 & 23.091162 & 0.4667\\
2 1 & 1 2 & 177.262010 & 0.641931 & 0.0056\\
2 1 & 1 1 & 177.263447 & 9.621011 & 0.0833\\
 & & & & \\
3 3 & 2 3 & 265.884887 & 9.277050 & 0.0370\\
3 2 & 2 1 & 265.886185 & 70.133541 & 0.2000\\
3 3 & 2 2 & 265.886431 & 74.216400 & 0.2963\\
3 4 & 2 3 & 265.886497 & 83.493759 & 0.4286\\
3 2 & 2 3 & 265.886976 & 0.371083 & 0.0011\\
3 2 & 2 2 & 265.888519 & 12.987986 & 0.0370\\
 & & & & \\
4 4 & 3 4 & 354.503893 & 12.826208 & 0.0208\\
4 3  & 3 2 & 354.505316 & 65.964149 & 0.2379\\
4 5 & 3 4 & 354.505458 & 51.305511 & 0.4084\\
4 4 & 3 3 & 354.505503 & 56.327704 & 0.3121\\
4 3 & 3 4 & 354.505841 & 1.832346 & 0.0003\\
4 3 & 3 3 & 354.507447 & 16.491334 & 0.0208\\
 & & & & \\
 5 5 & 4 5 & 443.114493 & 16.395434 & 0.0133\\
 5 4 & 4 3 & 443.116076 & 100.195394 & 0.2602\\
 5 6 & 4 5 & 443.116161 & 81.978097 & 0.3935\\
 5 5 & 4 4 & 443.116194 & 88.799114 & 0.3195\\
 5 4 & 4 5 & 443.116399 & 1.821738 & 0.0001\\
 5 4 & 4 4 & 443.118063 & 20.039348 & 0.0133\\ 
\bottomrule
\end{tabular}
\begin{tablenotes}
\item[] 
\end{tablenotes}
\end{threeparttable}
\label{tab:molspec}
\end{center}
\end{table}

The general Einstein A formulation for the dipole moment matrix coefficients $\rm\mu_{i=x,y,z}$ is given by:
\begin{equation}
\rm A_{l\leftarrow u} = \frac{64\pi^4\nu_{lu}^3}{3hc^3}[\left|\left<l|\mu_{x}|u\right>\right|^2 + \left|\left<l|\mu_{y}|u\right>\right|^2 + \left|\left<l|\mu_{z}|u\right>|^2\right].
\label{eq:qmeinstA}
\end{equation}
In order to arrive at an expression particular to hyperfine transitions, it is useful to express the dipole moment matrix coefficients according to a specific set of quantum mechanical eigenfunctions. For HCN, we chose the basis set given by $\rm\left|JIF\right>$, with vector addition of three angular momenta, the rotational angular momentum, J; the nuclear spin angular momentum, I; and the total angular momentum inclusive of spin, F; where {\bf F} = {\bf J}+{\bf I}. Considering hyperfine transitions therefore, the Einstein A becomes

\begin{equation}
A_{{\rm F\rightarrow F^{\prime}}} = \frac{64\pi^4\nu^3}{3hc^3}\|\mu\|^2\frac{g_{\rm tot}}{g_{\rm F}}\times S_{\rm F\rightarrow F^{\prime}},
\label{eq:einstA}
\end{equation}

\noindent
where, $\|\mu\|^2$=$\mu^2$J/(2J+1), $\mu$=2.984~debyes (D), and the line strength, S, for linear molecules is given by J/(2J+1). The ratio of the total degeneracy of the rotational level containing hyperfine structure to the degeneracy of the upper hyperfine state for the transition is g$\rm_{tot}$/g$\rm_F$. The normalised relative intensity of the transitions, S$_{\rm F\rightarrow F^{\prime}}$, is the intensity of the hyperfine transition compared to the total intensity of the parent rotational transition. It is given by:

\begin{equation}
S_{\rm F\rightarrow F^{\prime}} = \frac{(2F+1)(2F^{\prime}+1)}{2I+1}
\begin{Bmatrix}
J & F & I\\
F^{\prime} & J^{\prime} & 1\\
\end{Bmatrix}^2,
\label{eq:relinthyp}
\end{equation}

\noindent
where $\{:::\}$ is a Wigner 6-j symbol \citep[e.g., see][Table 5]{AngMomentQM}.

Putting Eqs.\eqref{eq:einstA}-\eqref{eq:relinthyp} together and simplifying gives the Einstein A coefficient for a hyperfine transition F$\rightarrow$F$^{\prime}$ of a linear molecule at frequency $\nu$ as,

\begin{equation}
A_{\rm F\rightarrow F^{\prime}} = \frac{64\pi^4\nu^3\mu^2}{3hc^3}(J)(2F^{\prime}+1)
\begin{Bmatrix}
J & F & I\\
F^{\prime} & J^{\prime} & 1\\
\end{Bmatrix}^2.
\label{eq:simpleA}
\end{equation}

In order to account for effects such as masering the Einstein B coefficients must also be known, and these can be calculated simply from the values of the Einstein A coefficients by considering detailed balance at thermal equilibrium.

The fourth and fifth columns of Table~\ref{tab:molspec} list, respectively, the Einstein A value, and the expected normalized relative intensity for each of the hyperfine transitions. These are given for the five lowest rotational transitions of HCN, the hyperfine lines become blended into one central line for transitions between higher rotational levels.

\subsection{Collisional Coefficient Formalism}

Radiative rates are readily available and independent of the environment in which the molecule is found (e.g. Eq.~\ref{eq:simpleA}). Collisional rates, however, are formed from the product of collisional rate coefficients and the population density of the collision partners. Ideally, these should be calculated exactly for HCN colliding with an appropriate mixture of ortho- and para-H$\rm _2$, Helium gas, and perhaps, electrons. However, the potential energy surface (PES) of H$_2$ is complex and requires intensive calculation. HCN collisional rates with  the much simpler spherical PES of He have been calculated for transitions between rotational levels.  These rates are then scaled to approximate the mixture of hydrogen and helium, firstly by \citet{GT74}, followed by \citet{Monteiro86}, and more recently by \citet{Dum10}. In recent years, advancements in computational power have allowed for the calculation of the first direct sets of HCN-H$_2$ rates. In \citet{BA12} the calculations considered H$_{\rm 2}$ as a structureless collisional partner. The results of this approximation differ from those obtained by \citet{Vera14}, who use a more advanced HCN-H$_{\rm 2}$ PES, taking fully into account the orientation of the H$_{\rm 2}$ molecule, performing scattering calculations which consider the rotational structure of both species, but the computational burden of the more precise calculations did not allow for the calculation of the individual F-level rate coefficients. Work is ongoing and hyperfine rates from the new PES are expected in the future. Electron collisional rates have been calculated \citep{Faure07} but will be negligible for these cloud conditions. 

For completeness we have investigated the effect of different possible collisional rates on the calculations to be described in the rest of this paper. The three main choices are:
\begin{itemize}
\item Directly calculated F-level rates \citep{BA12}.
\item The Proportional Method applied to J-level rates \citep{GT74,Monteiro86,Dum10,Vera14}.
\item Direct scaling of the \citet{Vera14} J-level rates using the F-level proportions calculated by \citet{BA12}. 
\end{itemize}

\subsubsection{The Proportional Method}
For such rates with only rotational J-levels given \citep{GT74,Monteiro86, Dum10, Vera14}, it is possible to use the `proportional method' to calculate collisional rate coefficients for all possible F-level hyperfine transitions, as initially suggested by \citet{Guill81}, and further demonstrated by \citet{Keto10}.  In the proportional method, the rates between each of the individual hyperfine levels are approximated as fractions of the net rotational collisional rate, scaled to the statistical degeneracy of the final hyperfine level. As the degeneracies of the hyperfine lines for a given rotational transition are different, the corresponding collisional coefficients will also be different. \citet{Stutzki85} suggested that this could be a possible underlying trait of the anomalies with the hyperfine lines excited selectively based on their collisional coefficients.  Thus the average net rates $\rm R_{J\rightarrow J^{\prime}}$ between rotational states: J,$\rm J^{\prime}$, are specified by way of a weighted sum over the hyperfine rate coefficients,
\begin{equation}
\rm C_{J\rightarrow J^{\prime}} = \displaystyle\sum_{FF^{\prime}}\frac{2F+1}{3(2J+1)}R_{JF\rightarrow J^{\prime}F^{\prime}},
\label{eq:rotrate}
\end{equation}
where we have implicitly assumed that the initial hyperfine states are occupied in proportion to their statistical weights. It follows that in the absence of the individual hyperfine collisional rate coefficients, the coefficients may be approximated assuming that each rate is proportional to the statistical weight of the final level \citep{Keto10}. In the case of HCN, these approximations for the rate coefficients take the form,
\begin{equation}
\rm C_{JF\rightarrow J^{\prime}F^{\prime}} = \frac{2F^{\prime}+1}{3(2J^{\prime}+1)}R_{JJ^{\prime}},
\label{eq:hyprate}
\end{equation}  
where the factor of 3 in the denominator comes from the statistical degeneracy of the nuclear spin quantum number (I~=~1), giving (2I+1)~=~3. We use rotational collisional rate coefficients C($\rm J\rightarrow J^{\prime}$) from \citet{GT74}, extrapolated and updated, as presented in \citet{Schoier2005}.  These are HCN-He rates and so must be scaled by a factor of 1.37 for collisions with para-H$_2$ .  

The rate coefficients scaled in this way can be completed with the inclusion of quasi-elastic rates where the rotational level does not change: $\Delta \rm{J} =0$, $\Delta \rm{F} \neq 0$.
These quasi-elastic rates were acquired based on an empirical approximation for the rotational rate coefficients made by \citet[Eq.(17)][]{deJong75}, by following the methodology devised by \citet[][see Section 4]{Keto10} as applied to N$_2$H$^+$.
\begin{equation}
\begin{split}
\rm C(J\rightarrow J^\prime) = \rm a\left( \Delta J\right) \frac{2J^\prime +1}{2J+1}\left( 1+\frac{\Delta E_{JJ^{\prime}}}{kT}\right)\\ \times\rm\exp\left[-b(\Delta J)\left(\frac{\Delta E_{JJ^{\prime}}}{kT}\right)^{\sfrac{1}{2}}\right]. 
\label{eq:deJong}
\end{split}
\end{equation}

For each of the 14 temperatures analysed between T=5 and $1200~{\rm K}$ inclusive, the fitting function, Eq.~\eqref{eq:deJong}, was manipulated so that the fitting variables a($\rm\Delta J$) and b($\rm\Delta J$) could be evaluated:
\begin{align}
\begin{split}
\hspace{2em}&\hspace{-2em} \rm\ln\left[C(J\rightarrow J^\prime)\frac{g_J}{g_{J^{\prime}}}\frac{1}{1+\sfrac{\Delta E_{JJ^{\prime}}}{kT}}\right]\\
&= \rm\ln\left[a\left(\Delta J\right)\right] \rm-b\left(\Delta J\right)\left(\frac{\Delta E_{JJ^{\prime}}}{kT}\right)^{\sfrac{1}{2}}.
\label{eq:InvdeJong}
\end{split}
\end{align}

\begin{figure*}
\begin{centering}
\includegraphics[width=0.7\textwidth]{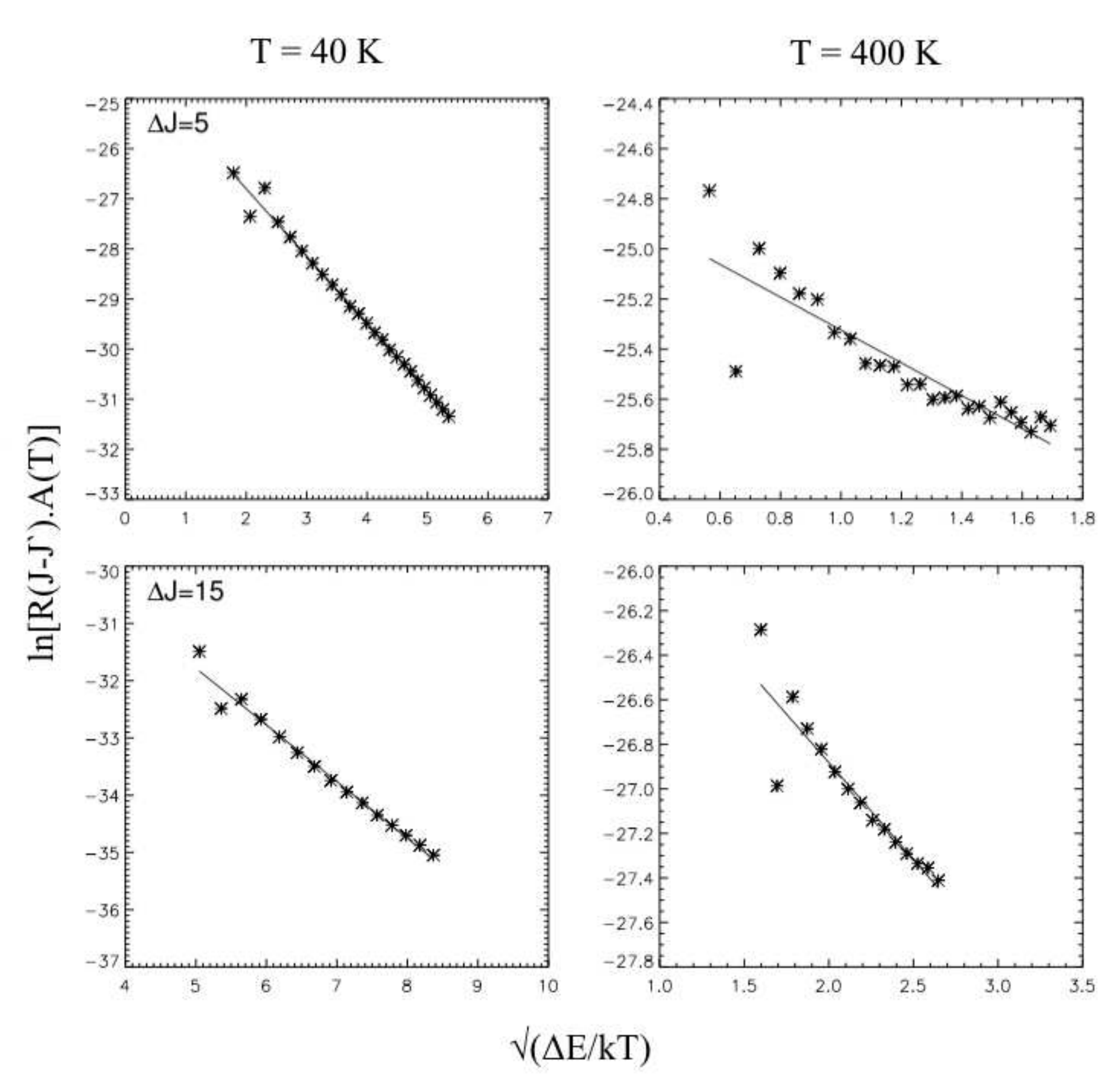}
\caption{The linear plot of Eq~\ref{eq:InvdeJong} for $\Delta$J=5 and $\Delta$J=15, respectively, at T=40K (\textit{left}) and T=400K (\textit{right}), where on the y-axis, A(T) = $(g_J/g_{J^{\prime}})\times1/[1+(\sfrac{\Delta E_{JJ^{\prime}}}{kT})]$}.
\label{quasiplots}
\end{centering}
\end{figure*}

The resultant equation, Eq.~\eqref{eq:InvdeJong}, is a straight line, y=mx+c. A set of plots was obtained for each value of $\rm\Delta J$ by plotting the LHS of Eq.~\eqref{eq:InvdeJong} against $\left(\sfrac{\Delta E_{JJ^{\prime}}}{kT}\right)^{1/2}$ (see Figure~\ref{quasiplots}). A least squares linear fit was performed where the coefficients b($\Delta$J) and $\ln\left[a(\rm\Delta J)\right]$ are given by the slope and y-intercept of this fit, respectively. 

\begin{figure*}
\begin{centering}
\includegraphics[width=0.7\textwidth]{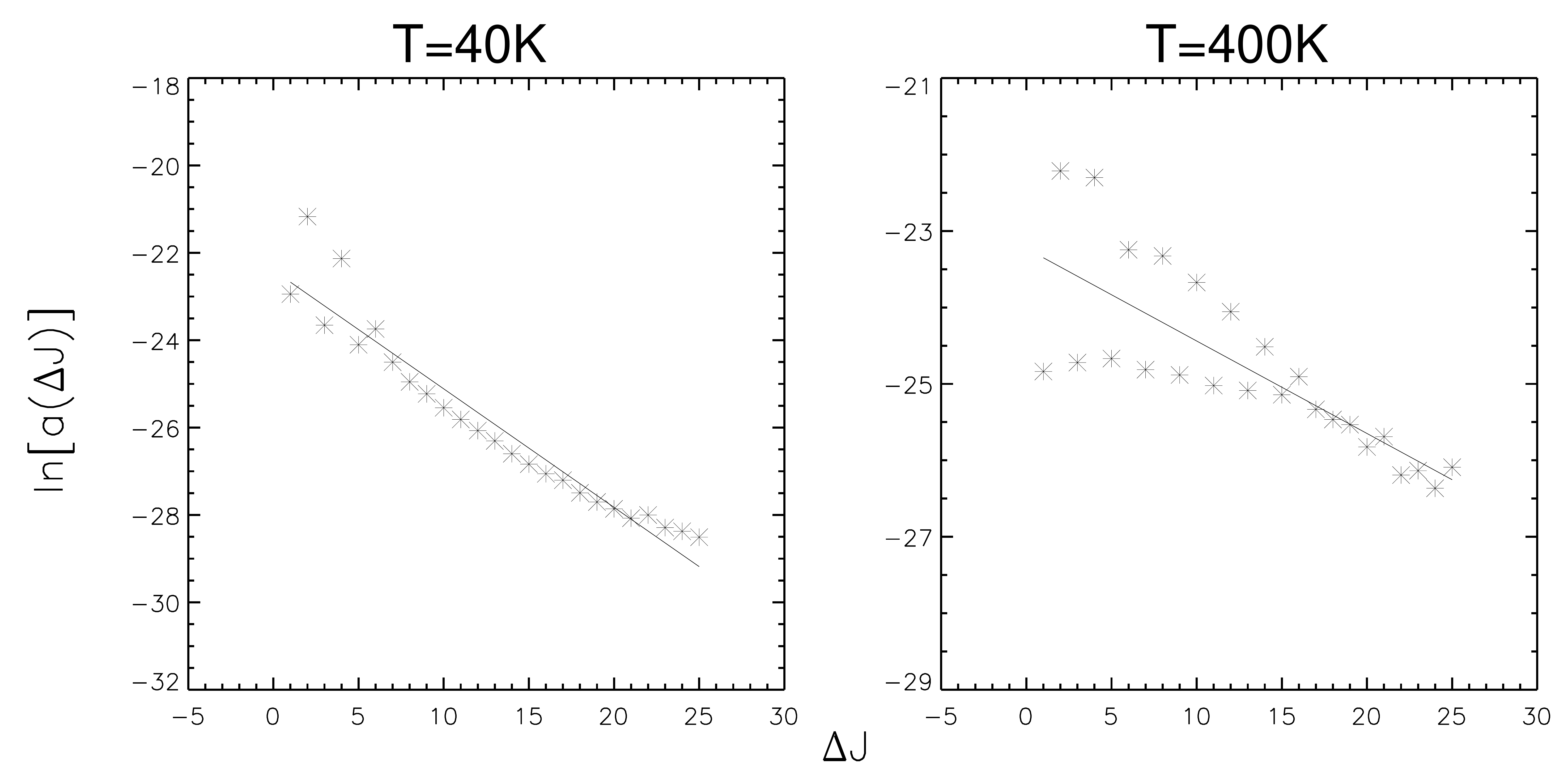}
\caption{Interpolation of fitting coefficients for ln[a($\rm\Delta J$)] at T=40K (\textit{left}) and T=400K (\textit{right}). The linear interpolation was carried out in order to deduce the values of ln[a($\rm\Delta J$)] for $\rm\Delta J$=0. As alluded to in the text, there is a clear shift in the fitting coefficient values for lower $\Delta$J whereby there is a strong preference for even $\Delta$J-transitions reflecting the obvious propensity rule for such transitions with this particular species.}
\label{fig:abfitting}
\end{centering}
\end{figure*}

Once the list of a and b fitting coefficients were collected for a particular temperature and $\rm\Delta J$, each was plotted against $\rm\Delta J$ so that an interpolation of the values present could be performed. The y-intercept following such an interpolation, as demonstrated in Figure~\ref{fig:abfitting}, translates to the fitting coefficient corresponding to the quasi-elastic rotational transition for a particular temperature (i.e., $\rm\Delta J$=0). The upper two panels in Figure~\ref{fig:abfitting} show that the collisional propensity rule is apparent for low $\Delta$J from the bifurcation of the data values into two streams; the bifurcating pattern tends to increase as the gas temperature rises. Due to the simplicity of the de Jong formula, most of its quantities diminish to zero upon the inclusion of quasi-elastic transitions. The rate coefficient for a particular quasi-elastic rotational transition at a given temperature reduces to the value for the fitting coefficient, $\rm a\left(\Delta J=0\right)$. The values of these coefficients are given in Table~\ref{quasitable}.

\subsection{Direct Scaling of \citet{Vera14} rates}
For the rates of \citet{Vera14}, although they are calculated at the J-level only, they consider also the structure of the H$_2$ colliding partner rather than taking angle averaged values when calculating their PES. This makes the calculation of exact F-level rates much more demanding. However, an approximation can be made by combining the two sets of rates together. Treating the F-level splitting as a perturbation on the J-level calculation, we generate a set of F-level rates based on the rates of \citet{Vera14} by taking the parent J-level coefficients and scaling them into the same proportions as the F-level rates calculated by \citet{BA12},  while maintaining the relation  $$\Sigma_{F'} k_{JF-J'F'}(T) = k_{J-J'}(T)$$ where $k_{JF-J'F'}(T)$ are the hyperfine rate coefficients, and $k_{J-J'}(T)$ are the rotational rate coefficients \citep{Fau12}. Table~\ref{vera-scale} gives an example of this calculation.\\

\begin{table}
\centering
\caption{An example of the scaling of the \citet{Vera14} rate coefficients, for the hyperfine states in the J = 2$\to$1 transition. $i$ and $j$ represent the initial and final hyperfine state respectively, and bA Rates are the rate coefficients of \citet{BA12}. All rate coefficients are in units of cm$^3$~s$^{-1}$.}
\label{vera-scale}
\begin{tabular}{@{}cccccc@{}}
\toprule
\multicolumn{1}{l}{$i$} & \multicolumn{1}{l}{$j$} & \multicolumn{1}{l}{bA Rates} & \multicolumn{1}{l}{Proportions} & \multicolumn{1}{l}{Vera Rate} & \multicolumn{1}{l}{Scaled Rates} \\ \midrule
5                                 & 2                               & 4.81E-12                           & 0.683529                        & 2.20E-11                             & 1.50E-11                               \\
5                                 & 3                               & 1.64E-12                           & 0.233055                        & 2.20E-11                             & 5.13E-12                               \\
5                                 & 4                               & 5.85E-13                           & 0.083133                        & 2.20E-11                             & 1.83E-12                               \\
6                                 & 2                               & 3.25E-12                           & 0.461844                        & 2.20E-11                             & 1.02E-11                               \\
6                                 & 3                               & 3.32E-12                           & 0.471792                        & 2.20E-11                             & 1.04E-11                               \\
6                                 & 4                               & 4.66E-13                           & 0.066222                        & 2.20E-11                             & 1.46E-12                               \\
7                                 & 2                               & 2.90E-12                           & 0.412107                        & 2.20E-11                             & 9.06E-12                               \\
7                                 & 3                               & 2.37E-12                           & 0.336792                        & 2.20E-11                             & 7.41E-12                               \\
7                                 & 4                               & 1.77E-12                           & 0.251529                        & 2.20E-11                             & 5.52E-12                               \\ \bottomrule
\end{tabular}
\end{table}

We find that \citet{Vera14} rates, both by the proportional method, and by the direct scaling method can readily reproduce the observed hyperfine anomalies of the low mass core TMC-1. We find similar results for the updated rates of  \citet{GT74} on the LAMDA Database \citep{Schoier2005}, with the proportional method applied, and supplemented with our quasi-elastic rates. We do note, however, that the results of our parameter sweep are largely unaffected by the choice of rates, which supports our assertion that carrying out radiative transfer over each F-level individually is essential. 

We note also that the proportional method approach has been applied to the linear radical $\rm N_2H^+$ by \citet{Keto10}, and in that work they verify the validity of the proportional method in its ability to reproduce the non-LTE hyperfine intensities in the Taurus dark cloud, L1512. \citet{Keto10} simulated observational data towards the L1512 starless core using hyperfine collisional rate coefficients derived using the theoretically-based study of \citet{Daniel05} and contrasted the results with simulations using the statistically weighted coefficients.  Both methods produced results that were similar to the accuracy expected from observations. 

While acknowledging that forthcoming rates based on the PES presented in \citet{Vera14} will represent the foremost rates for the HCN molecule when published, we use the scaled version of their J-level rates in all the calculations presented below. We note, however, that the proportional method applied to either \citet{GT74} or \citet{Vera14} gives very similar results when implemented in our radiative transfer scheme. Further analysis of the different collisional rate coefficients across a range of astrophysical conditions would be very worthwhile, but is beyond the scope of this work since the central conclusion, that the individual F-levels must be calculated in order to reproduce the anomalous HCN spectra, remains sound, irrespective of the choice of collisional rate coefficients.

\section{Radiative Transfer Modeling of HCN anomalies}

We use the above radiative and collisional rates for the hyperfine lines of HCN to modify the fully parallelised 3D radiative transfer code \textsc{mollie} \citep{Keto04} to calculate the strength and shape of each individual hyperfine line, including line overlap. The input to \textsc{mollie} is divided into voxels (3D pixels) and there are five input parameters which need to be uniquely defined in each voxel: the number density of H$_2$, the gas temperature, the gas bulk velocity, the gas turbulent velocity, and the relative abundance of the molecular species of interest (with respect to H$_{\rm 2}$).  In order to calculate the level populations, the statistical equilibrium equations are solved using an accelerated lambda iteration \citep{Rybicki91} that reduces the radiative transfer equations to a series of linear problems that are solved quickly even in optically thick conditions. Ray tracing is then used to generate synthetic line profiles from the model cube from arbirtary viewing angles to compare with observed lines. The final output spectra are groupings of individual hyperfine lines for a respective rotational transition (see Figure~\ref{HCN}). \textsc{mollie} is fully benchmarked against the test problems described in \citet{vanZadelhoff}. Note that an alternative 1D or 2D non-LTE radiative transfer code \citep{LIME,Juvela97}, using the collisional and radiative rates above, should be able to reproduce the spherically symmetric model results we present here.

\subsection{Model of the prototypical low mass source TMC-1}
\begin{figure}
\begin{centering}
\includegraphics[width=\linewidth]{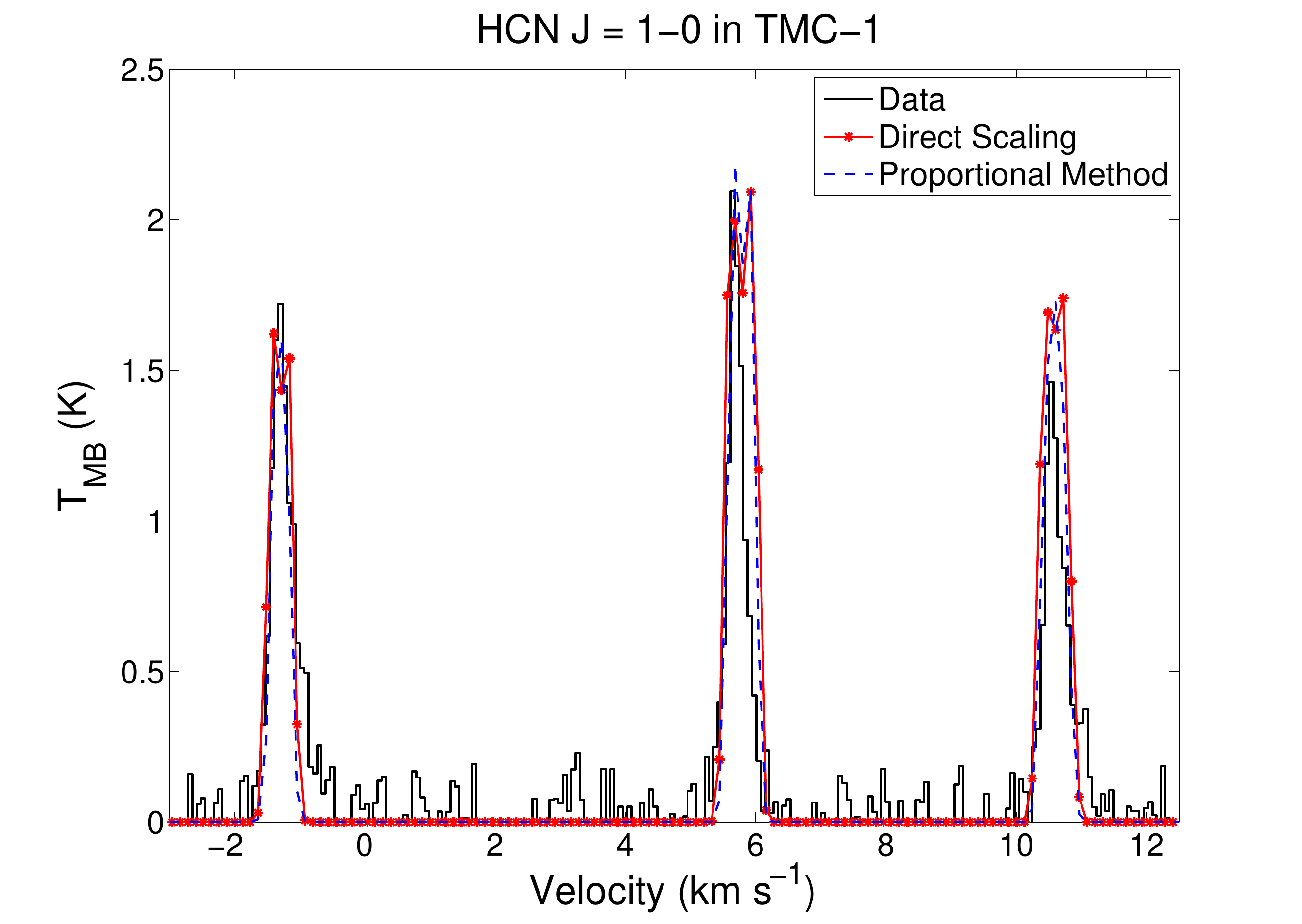}
\caption{Models of J=1$\to$0 line of HCN for the cloud TMC-1, using the scaled rates (dotted red line) and the proportional method (dashed blue line)  overlaid onto observed data (solid black line) from \citet{Sohn07}.}\label{TMC}
\end{centering}
\end{figure}

TMC-1 is a low mass star forming source in the Taurus Molecular Cloud, in which \citet{Walms82} first observed the hyperfine anomalies in HCN. We have fitted a simple model of the J=1$\to$0 transition of HCN in TMC-1, using observational data from \citet{Sohn07} observations. In Figure~\ref{TMC}, we show a good first order fit to the observed spectrum. In this model, we use as a starting point the densities, temperatures and abundances as measured by \citet{Pratap97}. Our model is a sphere of constant density, and temperature, with a low turbulent width, which is slowly infalling. It is notable straight away that the anomaly can be reproduced to first order, in a manner which is formally impossible in a radiative transfer treatment which considers only the rotational energy structure. We do not attempt a more detailed model of TMC-1 here, and note that more complex and realistic cloud dynamics can be implemented in low mass cores including rotation, outflows and freeze-out \citep{Carolan2008}. The simple model here is shown to illustrate that it should be readily possible to model the HCN line ratios in similar low mass sources, such as those seen in the data sets of \citet{Sohn07} and  \citet{Loughnane2012a},  provided the radiative transfer is carried out over each F-level individually.

\begin{table}
\caption{Physical Parameters for TMC-1 model. Errors indicate flexibility in choice of model parameters.  n$_{\rm H_2}$ is the number density of H$_{\rm 2}$ gas, T$_{\rm kin}$ is the kinetic temperature of the gas, v$_{\rm turb}$ is the micro turbulent velocity, v$_{\rm inf}$ is the bulk infall velocity, $\chi_{\rm HCN}$ is the relative abundance of HCN with respect to H$_{\rm 2}$. }\label{TMCtable}
\begin{center}
\begin{tabular}{lccc}
\hline\hline
Parameter & Prop. & Scaled & Error \\
\toprule
 n$_{\rm H_2}$ (cm$^{-3}$) & 6 $\times$ 10$^{4}$\ &6 $\times$ 10$^{4}$ & $\pm$ 0.5 $\times$ 10$^{4}$\\
T$_{\rm kin}$ (K) & 8 & 8  & $\pm$ 0.5\\
v$_{\rm turb}$ (cm s$^{-1}$) & 1.2 $\times$ 10$^{3}$& 1.2 $\times$ 10$^{3}$ & $\pm$ 0.2 $\times$ 10$^{3}$ \\
v$_{\rm inf}$ (cm s$^{-1}$) & 1 $\times$ 10$^{3}$ & 1 $\times$ 10$^{3}$ & $\pm$ 0.2$\times$ 10$^{3}$ \\
$\chi_{\rm HCN}$ (cm$^{-3}$) & 3 $\times$ 10$^{-9}$ & 2.75 $\times$ 10$^{-9}$ & $\pm$ 0.25 $\times$ 10$^{-9}$     \\  
\bottomrule
\end{tabular}
\end{center}
\end{table}

\subsection{Red-blue asymmetry switching in double-peaked line profiles}

\begin{figure*}
\begin{centering}
\includegraphics[width=0.45\linewidth]{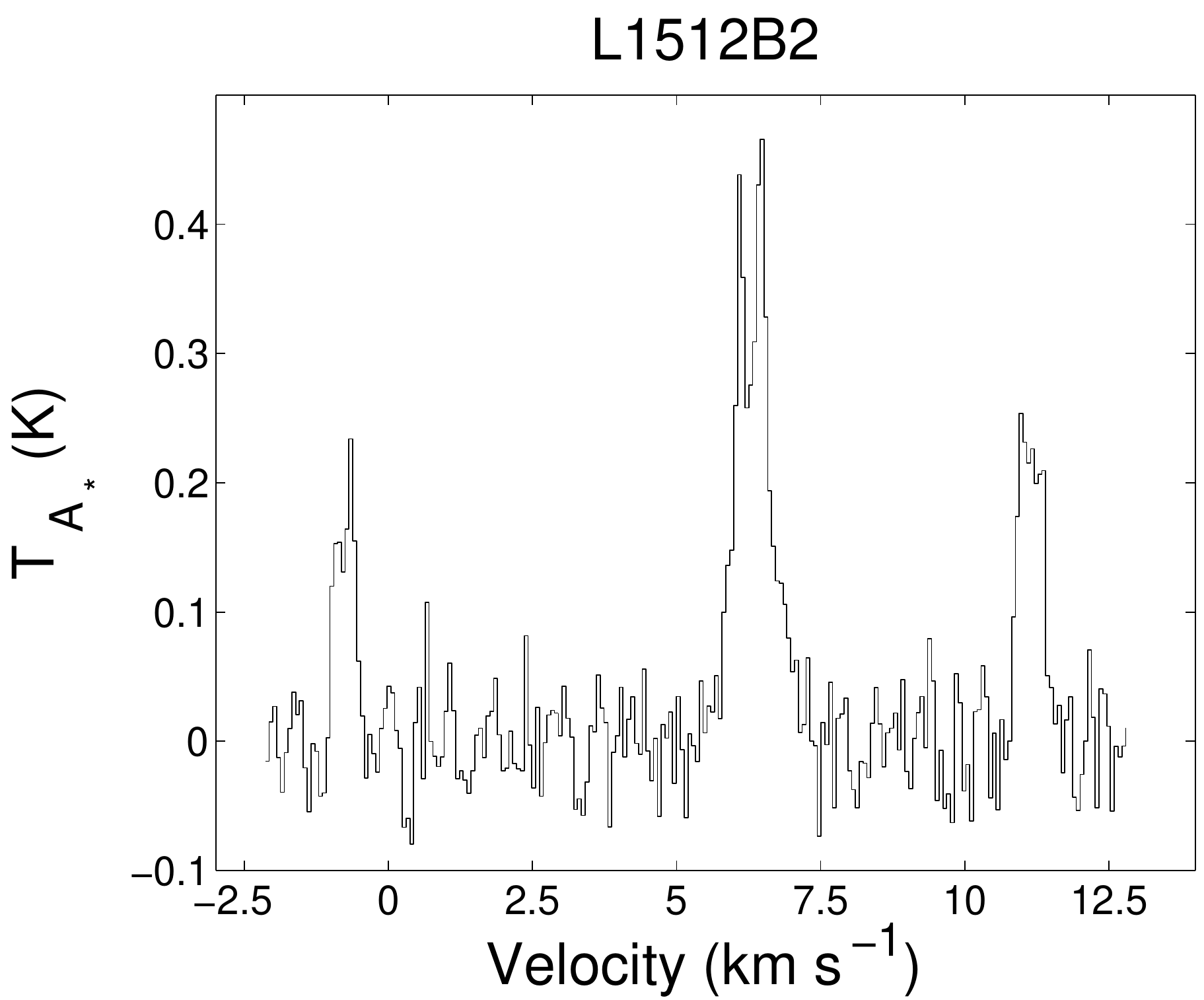}\qquad
\includegraphics[width=0.45\linewidth]{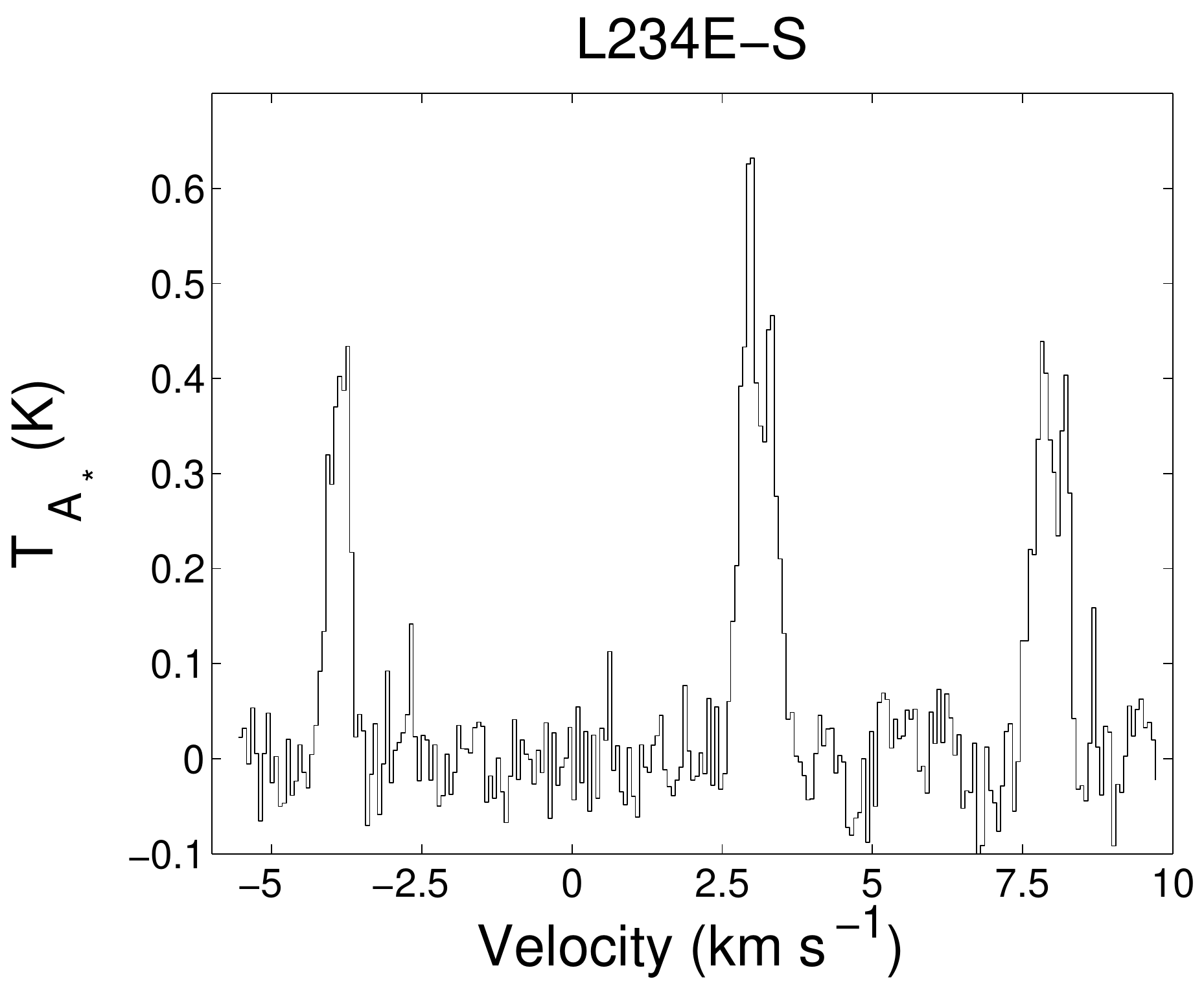}\\~\\
\includegraphics[width=0.45\linewidth]{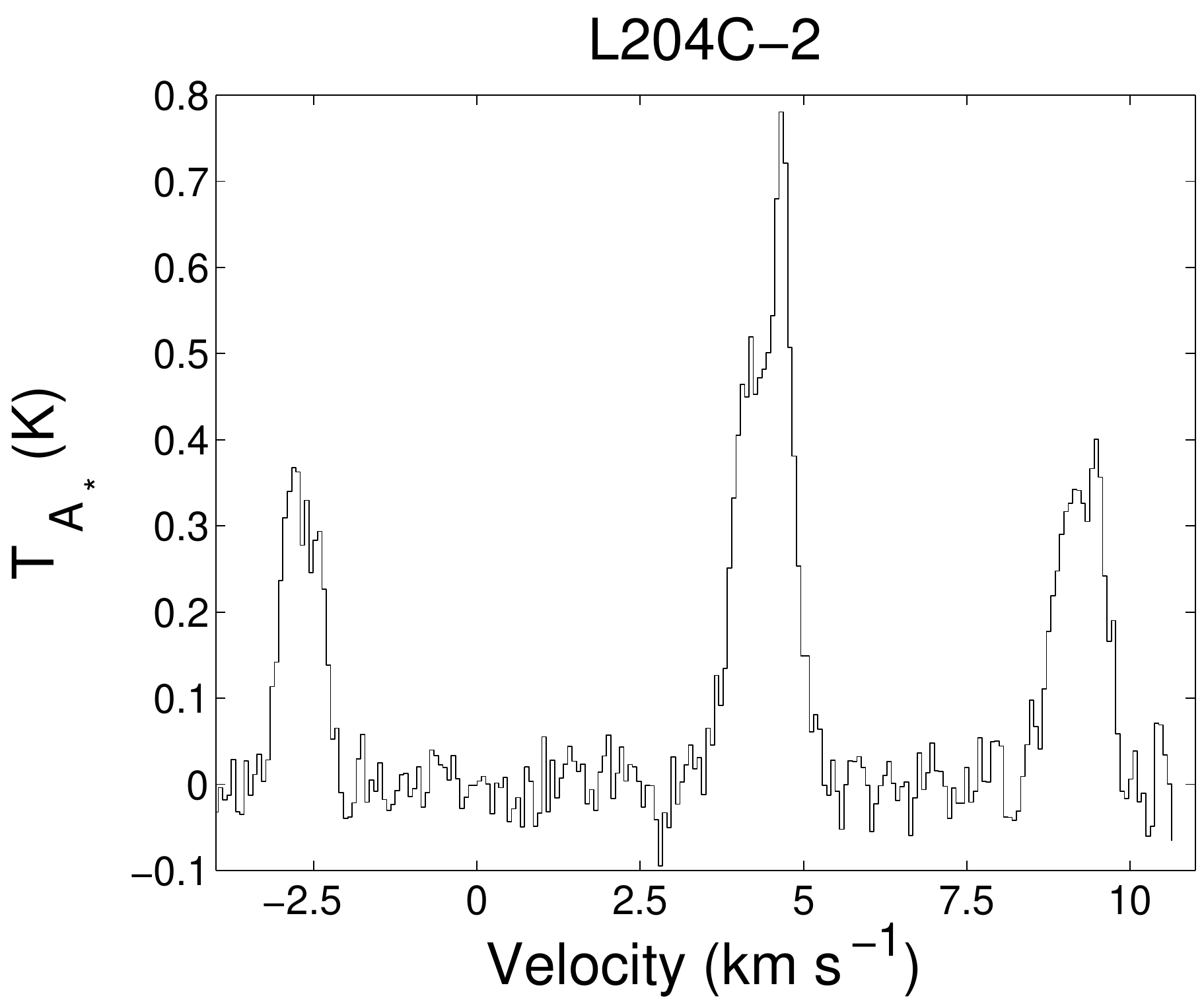}\qquad
\includegraphics[width=0.45\linewidth]{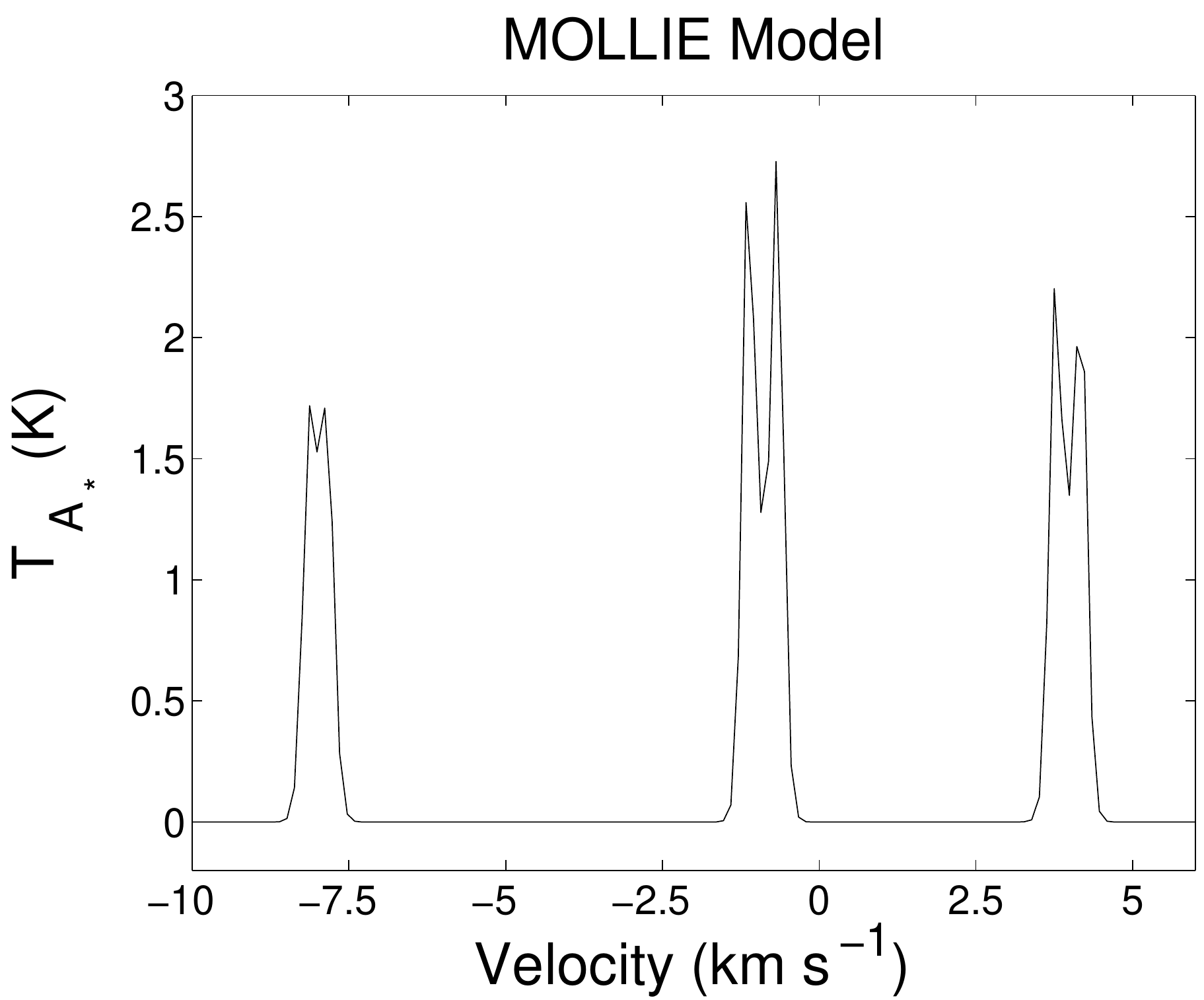}
\caption{The first three panels show example cores (L1512B2, L234E-S and L204C-2) exhibiting blue-red asymmetric switching in observational data \citep{Sohn07}. The final panel gives an example synthetic HCN J=1$\rightarrow$0 spectrum extracted from the parameter sweep of \S3.3, showing the capability of \textsc{mollie} in reproducing this signature.}\label{LRBsrc}
\end{centering}
\end{figure*}

Doppler shifting of gas components in a molecular cloud that is undergoing dynamical processes such as collapse, expansion or rotation can lead to double-peaked line profiles. In optically thick clouds, there can be an asymmetry in the strength of the peaks. In particular, for a collapsing cloud the blue peak of the line profile can be stronger than the red peak \citep{Evans99}. However, another peculiar aspect of the HCN spectrum concerns the red-blue asymmetry of some hyperfine lines such as those seen in Figure~\ref{LRBsrc}, and in L234E-S by \citet[see their Figure 1]{Schnee2013}. The asymmetries are observed to switch across the hyperfine lines within the $J=1\rightarrow 0$ level, with one line having the opposite asymmetry to the other two. Again, this is formally impossible to reproduce in an analysis of the HCN spectrum by rotational level only.
Using the {\sc mollie} HCN hyperfine treatment, an exploration of the parameter space reveals that this effect is due to an interplay between self-absorption and infall. For a model of constant parameters (velocity, density, turbulence, temperature, and abundance) and with a high optical thickness, even a slight infall velocity begins to present asymmetry switching which is only strengthened as the infall velocity is raised. The asymmetry switching can also be strengthened or weakened by increasing or reducing the degree of self-absorption respectively. This is achieved by altering the gas density or molecular abundance of the model. With further work this particular signature could be developed into a sensitive diagnostic tool.

\begin{table}
\begin{center}
\caption{Functional Form of Physical Parameters for Parameter Sweep. $r$ is the radial coordinate of the model, T$_{\rm kin}$, n and v$_{\rm inf}$ are the kinetic temperature, number density of H$_{\rm 2}$ and infall velocity respectively. }\label{func}
\begin{tabular}{lc}
\hline\hline
Parameter & Function\\
\toprule
Temperature & \scalebox{1.1}{T(r) = $\frac{T_{\rm kin}}{2}\left(\frac{r+1}{r}\right)$} \\[5pt]
Density & n(r) = \scalebox{1.1}{$\frac{ṇ_0}{5}\left(\frac{3r+2}{r}\right)$} \\[5pt]  
Velocity & v(r) = \scalebox{1.1}{v(x,y,z) = $v_{\rm inf}\sqrt{\frac{r_0}{r}}$}\\[5pt]
\bottomrule
\end{tabular}
\label{tab:physfuncs}
\end{center}
\end{table}

\begin{table*}
\begin{centering}
\caption{Sampling of Parameter Space.}\label{sweep}
\begin{threeparttable}
\begin{tabular}{llll}
\hline \hline
Parameter & Range & No. of Steps & Canonical Value\\
\toprule
Density (n$\rm_{H_2}$) & 1$\times$10$^4$~--~7$\times$10$^6~$cm$\rm^{-3}$ & 9 & 1$\times$10$^4~$cm$^{-3}$ \\
Temperature (T$\rm_{kin}$) & 6,~10,~15,~25,~40~K & 5 & 12~K\\
Abundance ($\rm\chi_{HCN}$) & 1$\times$10$^{-11}$~--~9$\times$10$^{-9}$ & 15& 1$\times$10$^{-11}$\\
Infall velocity (v$\rm_{inf}$) & 0,~0.1,~0.5,~1.0 $\times$ $\Delta$v & 4 & 0.3~kms$\rm^{-1}$\\
Turbulent width ($\Delta$v$\rm_{turb}$) & 0.1,~0.15 ,~0.2  kms$\rm^{-1}$  & 3 & 0.2~kms$\rm^{-1}$\\
\bottomrule
\end{tabular}
\begin{tablenotes}
\item[]  
\end{tablenotes}
\end{threeparttable}
\end{centering}
\end{table*}

\subsection{Parameter sweep across low-mass star forming conditions}
The \textsc{mollie} HCN hyperfine implementation seems to readily reproduce the HCN hyperfine spectrum of individual sources and is able to account for effects such as asymmetry switching. To use this treatment to widen the investigation of the anomalies, a parameter sweep through the 5 free parameters of the code was performed, to investigate the physical conditions giving rise to the hyperfine anomaly.  For each unique set of parameters, a simple model of a spherical cloud was run, with density, temperature and velocity profiles as outlined in Table~\ref{func}. The parameter space, and the sampling associated with each of the parameters, is outlined in Table~\ref{sweep}.


\begin{figure*}
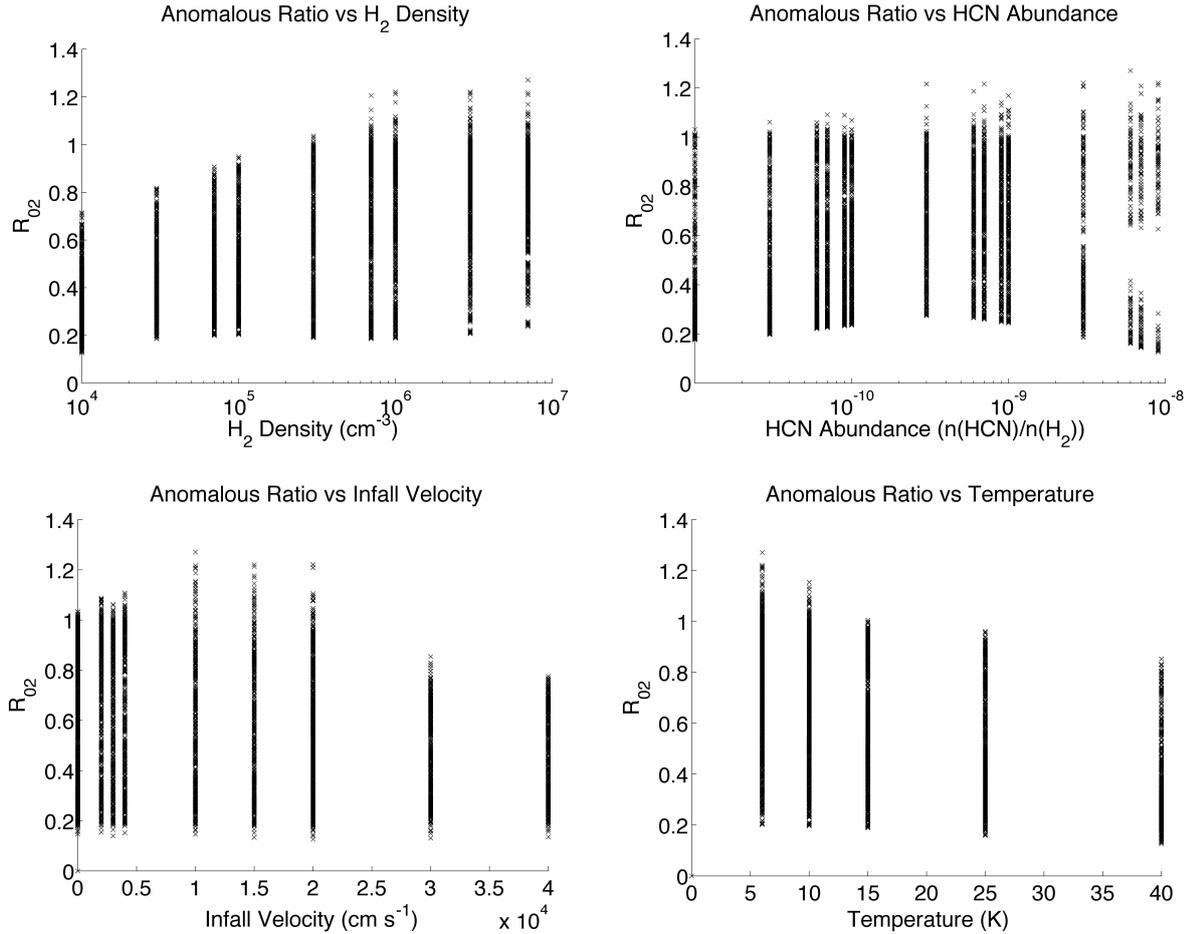
 
\begin{centering}
\includegraphics[width = 0.45\linewidth]{./HCN_Vera_Dens_Variation}
\includegraphics[width = 0.45\linewidth]{./HCN_Vera_Abun_Variation}\\~\\
\includegraphics[width = 0.45\linewidth]{./HCN_Vera_Inf_Variation}
\includegraphics[width = 0.45\linewidth]{./HCN_Vera_Temp_Variation}
\caption{Variation of R$_{02}$ with Density (\emph{ top left}), Abundance (\emph{top right}), Infall Velocity (\emph{bottom left}) and Temperature (\emph{bottom right}). Each small cross represents an individual model run.}\label{paravar}
\end{centering}
\end{figure*}

Once the parameter sweep was completed, we investigated its effect on the anomalous ratios, R$_{02}$ and R$_{12}$, as  each parameter was varied. Figure~\ref{paravar} shows the variation of R$_{02}$ as a function of each free parameter. There is a clear trend in the density plot showing that for a range of densities from $10^5$ to $10^6~{\rm cm^{-3}}$, any value of R$_{02}$ is possible, depending on the other parameters. Low densities tend towards the LTE optically thin case of 0.2, with some variation, while high densities tend towards the optically thick LTE value of 1, again with some variation. 
A bimodal distribution can be seen towards higher abundances, where the values tend towards the two LTE cases. This shows that the amount of HCN along the line of sight is the crucial factor, as expected, and also that the distribution tends towards the two LTE regimes, depending on the combination of the other parameters. The temperature displays a preference towards lower values for the most anomalous profiles.

Two important variables in the plots in Figure~\ref{paravar} were density and abundance. Conveniently, the product of these two variables is of major physical significance, it is the density of HCN in the cloud, which we will refer to as the HCN density ($\chi\,\rm{n}_{H_2}$). We can use this parameter to constrain the optical depth of the cloud and see how the anomalous ratios vary with it. The optical depth is given by $\tau \propto\,\chi\,n_{H_2}\Delta s \kappa_{\nu}$, where $\Delta$s  is the distance along the line of sight, $\chi n_{H_2}$ is the  absolute density of molecule of interest, and $\kappa_{\nu}$ is the opacity of a transition of frequency $\nu$. Since we have adopted a spherically symmetric model, with a constant line of sight for each run - and provided that the infall velocity is held constant, fixing the transition frequency, and thus the opacity - the value of the HCN density can then be used as a direct proxy for optical depth.   

\begin{figure}
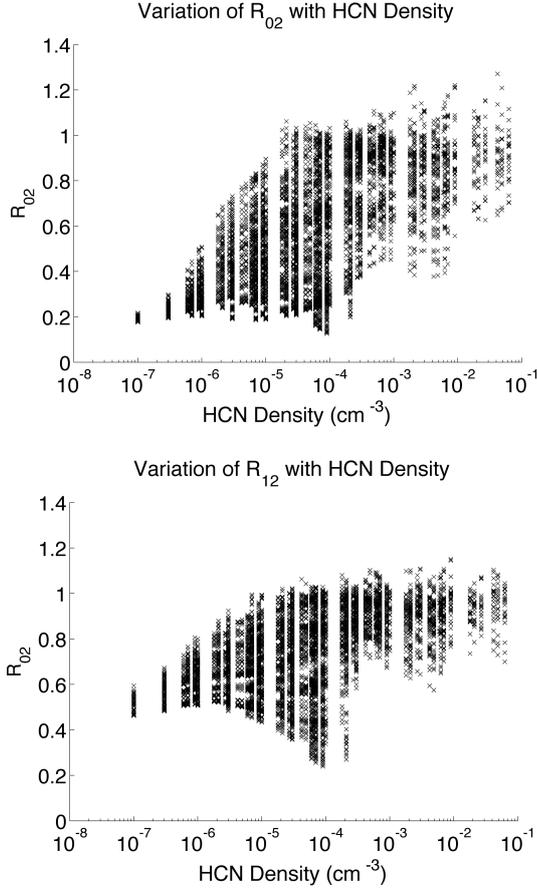
 
\begin{centering}
\includegraphics[width = 0.9\linewidth]{./R02_HCN_vera}\\~\\
\includegraphics[width = 0.9\linewidth]{./R12_HCN_vera}
\caption{Variation of R$_{02}$ (\emph{top}) and R$_{12}$(\emph{bottom}) with HCN density, $n_{\rm H_2}.\chi({\rm HCN})$, for an infall velocity of 2.0 km s$^{-1}$. Each small cross represents a separate model run.}\label{r12var}
\end{centering}
\end{figure}

The plots in Figure~\ref{r12var} show the possible values of  R$_{02}$ and R$_{12}$ as a function of HCN density for a unique value of the infall velocity in the parameter sweep. What we see, is that for low values of the HCN density, the values of R$_{02}$ and R$_{12}$ are in line with the expected LTE values of 0.2 and 0.6 respectively. As the HCN density (and thereby the optical depth) increases, we begin to see a range of possible values emerge. Using R$_{02}$ as an example, in the static cloud case, when $n_{\rm{HCN}} $ = 1$\times$10$^{-6}$ cm$^{-3}$, values range from 0.2 - 0.6, while if the density is  7$\times$10$^{-5}$ cm$^{-3}$  any value between 0.2 and 1.1 is possible. Similar trends can be seen for R$_{\rm{12}}$.  Also worth noting is the sharp transition at $n_{\rm{HCN}}$ = 1 $\times$ 10$^{-4}$ which is visible in all plots. This shows the transition from optically thin to optically thick, above which the anomalies tend towards the LTE value of 1. These plots clearly highlight the importance of the optical depth on the observed ratios. It should be noted that the values of R$_{\rm 02}$ which are much greater than 1, are readily reproduced by using the scaled rates, and are not reproduced by the proportional method.

\section{Conclusions}
We have demonstrated the anomalous behaviour of the J= 1$\to$ 0 line of HCN, and replicated both the observed anomalies in TMC-1, and the red-blue asymmetry switching typical of sources such as L1512B2, L234E-S and L204C-2. These effects have emerged from simple first order models, through applying radiative transfer at the F-level. Through a wide ranging parameter sweep we have demonstrated that the strengths of the satellite lines of the J=1$\to$0 transition are highly variable to changes in optical depth, and cannot be relied on to infer physical properties.

Based on our investigations, we suggest the following guidelines for interpreting HCN observations and for including HCN in radiative transfer codes.
\begin{itemize}
\item The radiative transfer of HCN must be carried out over each F-level individually.
\item In the J=1$\to$0 transition, the F=0$\to$1 line is unreliable as a diagnostic of infall or dynamics. The central F=2$\to$1 line however is more robust and when it is resolved, it should be reliable as a dynamical tracer. 
\item The J=2$\to$1 line is not usually observed due to its frequency location. However, it has been previously detected using IRAM \citep{Daniel2013a} and will be detectable with ALMA band 5 observations. The hyperfine components of this line are neither widely separated enough, nor centrally concentrated enough to be easily interpreted, and the line pattern will then be distorted further by the anomalies. The example model line in Figure~\ref{HCN} can be compared with the observation of the B1b low mass core carried out by \citet[][see Figure 13]{Daniel2013a}, to illustrate this point. It may be useful for some calculations to compare the integrated intensity of this line with other transitions.
\item In the J=3$\to$2 line, the central component is actually four overlapping hyperfines, and these can sometimes present as overly large infall signatures, particularly if interpreted as a single peak. The F = $2 \to 3$ component is boosted to far above its expected relative intensity (see Table~\ref{tab:molspec}, col.~5), and contributes to the distorted line shape of the central component (see Figure~\ref{HCN}). In general, the $\Delta$F = -1 component appears boosted in rotational transition spectra when hyperfine anomalies are prevalent promoting the line overlap phenomenon as the principle cause of these anomalous intensities.
\item For massive star forming regions the problem is worse as this already blended central component of the J=3$\to$2 line may be further blended with the satellite line marked B in Figure~\ref{HCN}. This can also present as a double peaked profile, which not only should not be used as a measure of the infall velocity, but also has a misplaced centroid velocity \citep[see Figure 15 from][]{Carolan09}.
\end{itemize}
For the above reasons we strongly advise against the use of the J=3$\to$2 line of HCN as a dynamical tracer, and advise that caution be used with J=1$\to$0, using only the central component.\\ 

Finally, the work of \citet{Keto10}, upon which this analysis is based, shows that a similar F-level radiative transfer treatment is required for N$_2$H$^+$, so we expect that all end N-bearing species such as HC$_3$N and the cyanopolyynes, NO and NH$_3$ could be susceptible to hyperfine anomalies, and these represent a target for future work. The central positive conclusion is that it is readily possible to reproduce the anomalous HCN spectrum, if the radiative transfer is carried out over individual hyperfine lines. 
 
\section*{Acknowledgements}
AMM acknowledges the support of the Irish Research Council EMBARK fellowship scheme, under whose funding this work took place. RML acknowledges financial support from DGAPA-UNAM through a postdoctoral fellowship. We thank the referee for suggesting carrying out a comparison between different collisional rate coefficients, which has lead to an improved paper. We would also like to acknowledge useful discussions with S. Lizano, J. Sheahan,  A. Ginsburg, D. Thornton, J.M.C. Rawlings, F. Lique, J. Tennyson, S. Viti, and D.A. Williams.








\appendix
\section{Collisional Rate Coefficients}

\begin{table*}
\caption{Quasi-elastic ($\Delta$J = 0) collisional rates for HCN up to J=7, and T = 100~K. Units of 10$^{-10}$~cm$^3$s$^{-1}$. Here J is the rotational level, F is the initial hyperfine level, and F' is the final hyperfine level. There are no quasi-elastic transitions with $\Delta$F = 0. }
\label{quasitable}
\begin{tabular}{llllllllll}
 &  &  &  &  & \textbf{T(K)} &  &  &  &  \\ \hline
\multicolumn{1}{|l|}{\textbf{J}} & \multicolumn{1}{l|}{\textbf{F}} & \multicolumn{1}{l|}{\textbf{F'}} & \multicolumn{1}{c|}{\textbf{5}} & \multicolumn{1}{c|}{\textbf{10}} & \multicolumn{1}{c|}{\textbf{20}} & \multicolumn{1}{c|}{\textbf{30}} & \multicolumn{1}{c|}{\textbf{40}} & \multicolumn{1}{c|}{\textbf{60}} & \multicolumn{1}{c|}{\textbf{100}} \\ \hline
\multicolumn{1}{|l}{1} & 1 & \multicolumn{1}{l|}{0} & \multicolumn{1}{l|}{0.1217} & \multicolumn{1}{l|}{0.2404} & \multicolumn{1}{l|}{0.2359} & \multicolumn{1}{l|}{0.2092} & \multicolumn{1}{l|}{0.2078} & \multicolumn{1}{l|}{0.1903} & \multicolumn{1}{l|}{0.0847} \\
\multicolumn{1}{|l}{1} & 2 & \multicolumn{1}{l|}{0} & \multicolumn{1}{l|}{0.1217} & \multicolumn{1}{l|}{0.2404} & \multicolumn{1}{l|}{0.2359} & \multicolumn{1}{l|}{0.2092} & \multicolumn{1}{l|}{0.2078} & \multicolumn{1}{l|}{0.1903} & \multicolumn{1}{l|}{0.0847} \\
\multicolumn{1}{|l}{1} & 0 & \multicolumn{1}{l|}{1} & \multicolumn{1}{l|}{0.3650} & \multicolumn{1}{l|}{0.7213} & \multicolumn{1}{l|}{0.7077} & \multicolumn{1}{l|}{0.6277} & \multicolumn{1}{l|}{0.6233} & \multicolumn{1}{l|}{0.5708} & \multicolumn{1}{l|}{0.2542} \\
\multicolumn{1}{|l}{1} & 2 & \multicolumn{1}{l|}{1} & \multicolumn{1}{l|}{0.3650} & \multicolumn{1}{l|}{0.7213} & \multicolumn{1}{l|}{0.7077} & \multicolumn{1}{l|}{0.6277} & \multicolumn{1}{l|}{0.6233} & \multicolumn{1}{l|}{0.5708} & \multicolumn{1}{l|}{0.2542} \\
\multicolumn{1}{|l}{1} & 0 & \multicolumn{1}{l|}{2} & \multicolumn{1}{l|}{0.6084} & \multicolumn{1}{l|}{1.2021} & \multicolumn{1}{l|}{1.1795} & \multicolumn{1}{l|}{1.0461} & \multicolumn{1}{l|}{1.0388} & \multicolumn{1}{l|}{0.9513} & \multicolumn{1}{l|}{0.4236} \\
\multicolumn{1}{|l}{1} & 1 & \multicolumn{1}{l|}{2} & \multicolumn{1}{l|}{0.6084} & \multicolumn{1}{l|}{1.2021} & \multicolumn{1}{l|}{1.1795} & \multicolumn{1}{l|}{1.0461} & \multicolumn{1}{l|}{1.0388} & \multicolumn{1}{l|}{0.9513} & \multicolumn{1}{l|}{0.4236} \\ \hline
\multicolumn{1}{|l}{2} & 2 & \multicolumn{1}{l|}{1} & \multicolumn{1}{l|}{0.2190} & \multicolumn{1}{l|}{0.4328} & \multicolumn{1}{l|}{0.4246} & \multicolumn{1}{l|}{0.3766} & \multicolumn{1}{l|}{0.3739} & \multicolumn{1}{l|}{0.3425} & \multicolumn{1}{l|}{0.1525} \\
\multicolumn{1}{|l}{2} & 3 & \multicolumn{1}{l|}{1} & \multicolumn{1}{l|}{0.2190} & \multicolumn{1}{l|}{0.4328} & \multicolumn{1}{l|}{0.4246} & \multicolumn{1}{l|}{0.3766} & \multicolumn{1}{l|}{0.3739} & \multicolumn{1}{l|}{0.3425} & \multicolumn{1}{l|}{0.1525} \\
\multicolumn{1}{|l}{2} & 1 & \multicolumn{1}{l|}{2} & \multicolumn{1}{l|}{0.3650} & \multicolumn{1}{l|}{0.7213} & \multicolumn{1}{l|}{0.7077} & \multicolumn{1}{l|}{0.6277} & \multicolumn{1}{l|}{0.6233} & \multicolumn{1}{l|}{0.5708} & \multicolumn{1}{l|}{0.2542} \\
\multicolumn{1}{|l}{2} & 3 & \multicolumn{1}{l|}{2} & \multicolumn{1}{l|}{0.3650} & \multicolumn{1}{l|}{0.7213} & \multicolumn{1}{l|}{0.7077} & \multicolumn{1}{l|}{0.6277} & \multicolumn{1}{l|}{0.6233} & \multicolumn{1}{l|}{0.5708} & \multicolumn{1}{l|}{0.2542} \\
\multicolumn{1}{|l}{2} & 1 & \multicolumn{1}{l|}{3} & \multicolumn{1}{l|}{0.5111} & \multicolumn{1}{l|}{1.0098} & \multicolumn{1}{l|}{0.9908} & \multicolumn{1}{l|}{0.8782} & \multicolumn{1}{l|}{0.8726} & \multicolumn{1}{l|}{0.7991} & \multicolumn{1}{l|}{0.3558} \\
\multicolumn{1}{|l}{2} & 2 & \multicolumn{1}{l|}{3} & \multicolumn{1}{l|}{0.5111} & \multicolumn{1}{l|}{1.0098} & \multicolumn{1}{l|}{0.9908} & \multicolumn{1}{l|}{0.8782} & \multicolumn{1}{l|}{0.8726} & \multicolumn{1}{l|}{0.7991} & \multicolumn{1}{l|}{0.3558} \\ \hline
\multicolumn{1}{|l}{3} & 3 & \multicolumn{1}{l|}{2} & \multicolumn{1}{l|}{0.2607} & \multicolumn{1}{l|}{0.5152} & \multicolumn{1}{l|}{0.5055} & \multicolumn{1}{l|}{0.4483} & \multicolumn{1}{l|}{0.4452} & \multicolumn{1}{l|}{0.4077} & \multicolumn{1}{l|}{0.1816} \\
\multicolumn{1}{|l}{3} & 4 & \multicolumn{1}{l|}{2} & \multicolumn{1}{l|}{0.2607} & \multicolumn{1}{l|}{0.5152} & \multicolumn{1}{l|}{0.5055} & \multicolumn{1}{l|}{0.4483} & \multicolumn{1}{l|}{0.4452} & \multicolumn{1}{l|}{0.4077} & \multicolumn{1}{l|}{0.1816} \\
\multicolumn{1}{|l}{3} & 2 & \multicolumn{1}{l|}{3} & \multicolumn{1}{l|}{0.3650} & \multicolumn{1}{l|}{0.7213} & \multicolumn{1}{l|}{0.7077} & \multicolumn{1}{l|}{0.6277} & \multicolumn{1}{l|}{0.6233} & \multicolumn{1}{l|}{0.5708} & \multicolumn{1}{l|}{0.2542} \\
\multicolumn{1}{|l}{3} & 4 & \multicolumn{1}{l|}{3} & \multicolumn{1}{l|}{0.3650} & \multicolumn{1}{l|}{0.7213} & \multicolumn{1}{l|}{0.7077} & \multicolumn{1}{l|}{0.6277} & \multicolumn{1}{l|}{0.6233} & \multicolumn{1}{l|}{0.5708} & \multicolumn{1}{l|}{0.2542} \\
\multicolumn{1}{|l}{3} & 2 & \multicolumn{1}{l|}{4} & \multicolumn{1}{l|}{0.4693} & \multicolumn{1}{l|}{0.9273} & \multicolumn{1}{l|}{0.9099} & \multicolumn{1}{l|}{0.8070} & \multicolumn{1}{l|}{0.8014} & \multicolumn{1}{l|}{0.7339} & \multicolumn{1}{l|}{0.3268} \\
\multicolumn{1}{|l}{3} & 3 & \multicolumn{1}{l|}{4} & \multicolumn{1}{l|}{0.4693} & \multicolumn{1}{l|}{0.9273} & \multicolumn{1}{l|}{0.9099} & \multicolumn{1}{l|}{0.8070} & \multicolumn{1}{l|}{0.8014} & \multicolumn{1}{l|}{0.7339} & \multicolumn{1}{l|}{0.3268} \\ \hline
\multicolumn{1}{|l}{4} & 4 & \multicolumn{1}{l|}{3} & \multicolumn{1}{l|}{0.2839} & \multicolumn{1}{l|}{0.5610} & \multicolumn{1}{l|}{0.5504} & \multicolumn{1}{l|}{0.4882} & \multicolumn{1}{l|}{0.4848} & \multicolumn{1}{l|}{0.4439} & \multicolumn{1}{l|}{0.1977} \\
\multicolumn{1}{|l}{4} & 5 & \multicolumn{1}{l|}{3} & \multicolumn{1}{l|}{0.2839} & \multicolumn{1}{l|}{0.5610} & \multicolumn{1}{l|}{0.5504} & \multicolumn{1}{l|}{0.4882} & \multicolumn{1}{l|}{0.4848} & \multicolumn{1}{l|}{0.4439} & \multicolumn{1}{l|}{0.1977} \\
\multicolumn{1}{|l}{4} & 3 & \multicolumn{1}{l|}{4} & \multicolumn{1}{l|}{0.3650} & \multicolumn{1}{l|}{0.7213} & \multicolumn{1}{l|}{0.7077} & \multicolumn{1}{l|}{0.6277} & \multicolumn{1}{l|}{0.6233} & \multicolumn{1}{l|}{0.5708} & \multicolumn{1}{l|}{0.2542} \\
\multicolumn{1}{|l}{4} & 5 & \multicolumn{1}{l|}{4} & \multicolumn{1}{l|}{0.3650} & \multicolumn{1}{l|}{0.7213} & \multicolumn{1}{l|}{0.7077} & \multicolumn{1}{l|}{0.6277} & \multicolumn{1}{l|}{0.6233} & \multicolumn{1}{l|}{0.5708} & \multicolumn{1}{l|}{0.2542} \\
\multicolumn{1}{|l}{4} & 3 & \multicolumn{1}{l|}{5} & \multicolumn{1}{l|}{0.4462} & \multicolumn{1}{l|}{0.8815} & \multicolumn{1}{l|}{0.8649} & \multicolumn{1}{l|}{0.7671} & \multicolumn{1}{l|}{0.7618} & \multicolumn{1}{l|}{0.6976} & \multicolumn{1}{l|}{0.3107} \\
\multicolumn{1}{|l}{4} & 4 & \multicolumn{1}{l|}{5} & \multicolumn{1}{l|}{0.4462} & \multicolumn{1}{l|}{0.8815} & \multicolumn{1}{l|}{0.8649} & \multicolumn{1}{l|}{0.7671} & \multicolumn{1}{l|}{0.7618} & \multicolumn{1}{l|}{0.6976} & \multicolumn{1}{l|}{0.3107} \\ \hline
\multicolumn{1}{|l}{5} & 5 & \multicolumn{1}{l|}{4} & \multicolumn{1}{l|}{0.2987} & \multicolumn{1}{l|}{0.5901} & \multicolumn{1}{l|}{0.5790} & \multicolumn{1}{l|}{0.5135} & \multicolumn{1}{l|}{0.5099} & \multicolumn{1}{l|}{0.4670} & \multicolumn{1}{l|}{0.2080} \\
\multicolumn{1}{|l}{5} & 6 & \multicolumn{1}{l|}{4} & \multicolumn{1}{l|}{0.2987} & \multicolumn{1}{l|}{0.5901} & \multicolumn{1}{l|}{0.5790} & \multicolumn{1}{l|}{0.5135} & \multicolumn{1}{l|}{0.5099} & \multicolumn{1}{l|}{0.4670} & \multicolumn{1}{l|}{0.2080} \\
\multicolumn{1}{|l}{5} & 4 & \multicolumn{1}{l|}{5} & \multicolumn{1}{l|}{0.3650} & \multicolumn{1}{l|}{0.7213} & \multicolumn{1}{l|}{0.7077} & \multicolumn{1}{l|}{0.6277} & \multicolumn{1}{l|}{0.6233} & \multicolumn{1}{l|}{0.5708} & \multicolumn{1}{l|}{0.2542} \\
\multicolumn{1}{|l}{5} & 6 & \multicolumn{1}{l|}{5} & \multicolumn{1}{l|}{0.3650} & \multicolumn{1}{l|}{0.7213} & \multicolumn{1}{l|}{0.7077} & \multicolumn{1}{l|}{0.6277} & \multicolumn{1}{l|}{0.6233} & \multicolumn{1}{l|}{0.5708} & \multicolumn{1}{l|}{0.2542} \\
\multicolumn{1}{|l}{5} & 4 & \multicolumn{1}{l|}{6} & \multicolumn{1}{l|}{0.4314} & \multicolumn{1}{l|}{0.8524} & \multicolumn{1}{l|}{0.8364} & \multicolumn{1}{l|}{0.7418} & \multicolumn{1}{l|}{0.7366} & \multicolumn{1}{l|}{0.6746} & \multicolumn{1}{l|}{0.3004} \\
\multicolumn{1}{|l}{5} & 5 & \multicolumn{1}{l|}{6} & \multicolumn{1}{l|}{0.4314} & \multicolumn{1}{l|}{0.8524} & \multicolumn{1}{l|}{0.8364} & \multicolumn{1}{l|}{0.7418} & \multicolumn{1}{l|}{0.7366} & \multicolumn{1}{l|}{0.6746} & \multicolumn{1}{l|}{0.3004} \\ \hline
\multicolumn{1}{|l}{6} & 6 & \multicolumn{1}{l|}{5} & \multicolumn{1}{l|}{0.3089} & \multicolumn{1}{l|}{0.6103} & \multicolumn{1}{l|}{0.5988} & \multicolumn{1}{l|}{0.5311} & \multicolumn{1}{l|}{0.5274} & \multicolumn{1}{l|}{0.4830} & \multicolumn{1}{l|}{0.2151} \\
\multicolumn{1}{|l}{6} & 7 & \multicolumn{1}{l|}{5} & \multicolumn{1}{l|}{0.3089} & \multicolumn{1}{l|}{0.6103} & \multicolumn{1}{l|}{0.5988} & \multicolumn{1}{l|}{0.5311} & \multicolumn{1}{l|}{0.5274} & \multicolumn{1}{l|}{0.4830} & \multicolumn{1}{l|}{0.2151} \\
\multicolumn{1}{|l}{6} & 5 & \multicolumn{1}{l|}{6} & \multicolumn{1}{l|}{0.3650} & \multicolumn{1}{l|}{0.7213} & \multicolumn{1}{l|}{0.7077} & \multicolumn{1}{l|}{0.6277} & \multicolumn{1}{l|}{0.6233} & \multicolumn{1}{l|}{0.5708} & \multicolumn{1}{l|}{0.2542} \\
\multicolumn{1}{|l}{6} & 7 & \multicolumn{1}{l|}{6} & \multicolumn{1}{l|}{0.3650} & \multicolumn{1}{l|}{0.7213} & \multicolumn{1}{l|}{0.7077} & \multicolumn{1}{l|}{0.6277} & \multicolumn{1}{l|}{0.6233} & \multicolumn{1}{l|}{0.5708} & \multicolumn{1}{l|}{0.2542} \\
\multicolumn{1}{|l}{6} & 5 & \multicolumn{1}{l|}{7} & \multicolumn{1}{l|}{0.4212} & \multicolumn{1}{l|}{0.8322} & \multicolumn{1}{l|}{0.8166} & \multicolumn{1}{l|}{0.7242} & \multicolumn{1}{l|}{0.7192} & \multicolumn{1}{l|}{0.6586} & \multicolumn{1}{l|}{0.2933} \\
\multicolumn{1}{|l}{6} & 6 & \multicolumn{1}{l|}{7} & \multicolumn{1}{l|}{0.4212} & \multicolumn{1}{l|}{0.8322} & \multicolumn{1}{l|}{0.8166} & \multicolumn{1}{l|}{0.7242} & \multicolumn{1}{l|}{0.7192} & \multicolumn{1}{l|}{0.6586} & \multicolumn{1}{l|}{0.2933} \\ \hline
\multicolumn{1}{|l}{7} & 7 & \multicolumn{1}{l|}{6} & \multicolumn{1}{l|}{0.3164} & \multicolumn{1}{l|}{0.6251} & \multicolumn{1}{l|}{0.6133} & \multicolumn{1}{l|}{0.5440} & \multicolumn{1}{l|}{0.5402} & \multicolumn{1}{l|}{0.4947} & \multicolumn{1}{l|}{0.2203} \\
\multicolumn{1}{|l}{7} & 8 & \multicolumn{1}{l|}{6} & \multicolumn{1}{l|}{0.3164} & \multicolumn{1}{l|}{0.6251} & \multicolumn{1}{l|}{0.6133} & \multicolumn{1}{l|}{0.5440} & \multicolumn{1}{l|}{0.5402} & \multicolumn{1}{l|}{0.4947} & \multicolumn{1}{l|}{0.2203} \\
\multicolumn{1}{|l}{7} & 6 & \multicolumn{1}{l|}{7} & \multicolumn{1}{l|}{0.3650} & \multicolumn{1}{l|}{0.7213} & \multicolumn{1}{l|}{0.7077} & \multicolumn{1}{l|}{0.6277} & \multicolumn{1}{l|}{0.6233} & \multicolumn{1}{l|}{0.5708} & \multicolumn{1}{l|}{0.2542} \\
\multicolumn{1}{|l}{7} & 8 & \multicolumn{1}{l|}{7} & \multicolumn{1}{l|}{0.3650} & \multicolumn{1}{l|}{0.7213} & \multicolumn{1}{l|}{0.7077} & \multicolumn{1}{l|}{0.6277} & \multicolumn{1}{l|}{0.6233} & \multicolumn{1}{l|}{0.5708} & \multicolumn{1}{l|}{0.2542} \\
\multicolumn{1}{|l}{7} & 6 & \multicolumn{1}{l|}{8} & \multicolumn{1}{l|}{0.4317} & \multicolumn{1}{l|}{0.8174} & \multicolumn{1}{l|}{0.8020} & \multicolumn{1}{l|}{0.7113} & \multicolumn{1}{l|}{0.7064} & \multicolumn{1}{l|}{0.6469} & \multicolumn{1}{l|}{0.2881} \\
\multicolumn{1}{|l}{7} & 7 & \multicolumn{1}{l|}{8} & \multicolumn{1}{l|}{0.4317} & \multicolumn{1}{l|}{0.8174} & \multicolumn{1}{l|}{0.8020} & \multicolumn{1}{l|}{0.7113} & \multicolumn{1}{l|}{0.7064} & \multicolumn{1}{l|}{0.6469} & \multicolumn{1}{l|}{0.2881} \\ \hline
\end{tabular}
\end{table*}

\bsp	
\label{lastpage}
\end{document}